%% file: draft_02.tex
\documentclass[trackchanges, twocolumn, twocolappendix]{aastex701}
\input{def.tex}

\RenewDocumentCommand{\added}{O{} O{} m}{\textcolor{Black}{#3}}

\begin{document}

\title{X-ray reflection as diagnostic of supermassive black hole binary properties}

\author[0009-0004-2625-5527]{Julie Malewicz}
\affiliation{Center for Relativistic Astrophysics, School of Physics,
  Georgia Institute of Technology, 837 State Street NW, Atlanta, GA 30332-0430, USA}
\email[show]{jlm@gatech.edu}

\author[0000-0002-7835-7814]{Tamara Bogdanovi\'c}
\affiliation{Center for Relativistic Astrophysics, School of Physics,
  Georgia Institute of Technology, 837 State Street NW, Atlanta, GA 30332-0430, USA}
\email{tamarab@gatech.edu}

\author[0000-0001-8128-6976]{David R.\  Ballantyne}
\affiliation{Center for Relativistic Astrophysics, School of Physics,
  Georgia Institute of Technology, 837 State Street NW, Atlanta, GA 30332-0430, USA}
\email{david.ballantyne@gatech.edu}

\author[0000-0003-2663-1954]{Laura Brenneman}
\affiliation{Center for Astrophysics, Harvard-Smithsonian, 60 Garden Street, Cambridge, MA 02138, USA}
\email{lbrenneman@cfa.harvard.edu}

\author[0000-0003-4583-9048]{Thomas Dauser}
\affiliation{Dr. Karl Remeis-Observatory \& ECAP, FAU Erlangen-Nürnberg,
  Sternwartstr. 7, 96049 Bamberg, Germany}
\email{thomas.dauser@sternwarte.uni-erlangen.de}

\begin{abstract}
We investigate correlations between prominent features in the relativistic X-ray reflection spectrum emitted by an accreting supermassive black hole (SMBH) binary with the underlying properties of the binary system.
Model-independent measurements of the relativistic \feka\ line ($\sim$6.4 keV) and the Compton reflection hump ($\sim$20-30 keV) are shown to be useful in constraining binary parameters.
We compute 24,570 X-ray reflection spectra from two mini-disks attached to SMBHs at a fixed separation of 100~$GM/c^2$ on circular orbits, by varying mass ratio, spin parameters,  inclination, orbital phase and total mass accretion rate.
We find that the location of the blue peak in the relativistic \feka\ is a relatively robust diagnostic of the binary
inclination,
which could be obtained from a single-epoch X-ray spectrum.
Given a few epochs of spectra, one may be able to determine the orbital phase of the binary and place constraints on its mass ratio by monitoring the \feka\ centroid.
Of all parameters, SMBH spin effects are most subtle and prone to measurement degeneracies.
Some markers of high SMBH spin may nevertheless surface in the composite spectrum due to increased radiative efficiency.
The approach developed here can be used to place preliminary constraints on binary parameters before a full parametrized X-ray spectral fitting method is available.
It complements gravitational wave measurements by the Pulsar Timing Arrays (PTAs) and the Laser Interferometer Space Antenna by providing independent constraints on binary parameters that may be prone to degeneracy (inclination), or otherwise inaccessible (spin for PTAs).\\

\end{abstract}

\section{Introduction} 
Standard hierarchical models of galaxy evolution predict that following galaxy mergers, their two central supermassive black holes (SMBHs) may sink towards the nucleus of the newly formed galaxy through dynamical friction processes, eventually pairing up into a gravitationally bound binary \citep[e.g.,][]{begelman1980,hopkins2008}. Once the orbit reaches sub-parsec separation, gravitational wave emission then becomes the dominant process driving the binary towards merger \citep{peters1963}.
Galaxy mergers are also expected to funnel large quantities of gas towards the nuclear region, enabling the two SMBHs to shine as binary active galactic nuclei (AGN) \citep[e.g.,][]{barnes1996,mihos1996}.
This makes them key multimessenger targets for joint electromagnetic (EM) and gravitational wave (GW) observations \citep[e.g.,][]{derosa2019}.

The low-frequency gravitational waves produced by these SMBH binaries will be detectable in the near future both through space-based interferometry and the continued operation of Galactic-scale timing experiments.
The Laser Interferometer Space Antenna \citep[LISA;][]{amaro-seoane2017}, in the millihertz regime, will be sensitive to inspiraling $\sim10^5-10^7 \Msun$ binaries and is expected to capture up to a few tens of MBH merger events during its lifetime \citep{amaro-seoane2023}. Pulsar Timing Arrays \citep[PTAs;][]{verbiest2016}, conversely, leverage the incredible regularity of pulsars strewn about the galaxy to detect nanohertz GWs from the most massive ($> 10^8 \Msun$) and nearby ($z\lesssim2$) SMBH binaries.
Though the long periods probed by PTAs require decades-long observational baselines to confidently claim individual detections, international PTA collaborations have reported the detection of a stochastic gravitational wave background of astrophysical origin, consistent with the population of SMBH binaries we expect from galaxy evolution models \citep{agazie2023,xu2023,reardon2023,antoniadis2023}.

Efforts are underway to detect SMBH binaries using a variety of EM approaches \citep[see][for recent reviews]{bogdanovic2022,doraziocharisi2023} in both time and frequency domains.
For example,
variable or Doppler-shifted optical broad lines consistent with binary orbital motion have been used to identify SMBH binary candidates at large orbital separations \citep[e.g.,][]{gaskell1996, eracleous2012,runnoe2017}. 
However, this diagnostic is only applicable over a limited range of orbital separations: wide enough to conserve at least one broad line region, yet tight enough for the kinematic shift to be measurable \citep[e.g.,][]{kelley2020}.
Accessing later stages of binary orbital evolution requires spectral diagnostics anchored closer to the SMBHs -- precisely the regime probed by X-ray spectroscopy.
Indeed, the bulk of the X-ray flux in AGN originates in the inner accretion disk, within a few tens of \rg\ (gravitational radii; $GM/c^2$) of the event horizon itself \citep[e.g.,][]{fabian1989}.
X-ray reflection spectroscopy probes this region directly by characterizing the ionized spectrum from the innermost disk under illumination by a compact corona \citep[e.g.,][]{garcia2010,dauser2013,ballantyne2017}. The corona -- a hot, optically thin electron plasma, is responsible for the up-scattering of thermal UV radiation into a hard X-ray power-law energy distribution via inverse Compton.
The emergent spectrum is then subject to significant gravitational and Doppler shifting effects, potentially allowing one to extract information about key SMBH properties, such as spin \citep[e.g.,][]{brenneman2006,dauser2013, reynolds2013, reynolds2021}. 

Extensive hydrodynamical simulations have laid the groundwork to expand these methods to binaries by mapping critical differences in their accretion dynamics compared to isolated SMBHs. These numerical simulations show that the orbital motion of the two SMBHs carves out a low-density cavity of diameter approximately twice their orbital separation.
Within this cavity, each SMBH retains its own individual, tidally truncated `mini-disk' \citep{papaloizou1977, lin1979, pichardo2005}, continuously resupplied from the surrounding circumbinary disk (CBD) through accretion streams
\citep[e.g.,][]{bowen2017,bowen2018,ryanmacfadyen2017,tang2018,dascoli2018,combi2022,avara2024,chan2025}.

\citet[][hereafter referred to as \citetalias{malewicz2025}]{malewicz2025} laid out a model to compute the composite relativistic ionized reflection spectrum from the 2 mini-disks tethered to SMBHs in a tight orbit (100~$GM/c^2\equiv 100~\rg$, where $M=M_1+M_2$ is the mass of the binary) around their common center of mass. 
We show that SMBH binaries are expected to produce distinctive composite spectra, with a characteristic time variability due to Doppler shifting induced by the SMBHs' orbital motion. In unequal mass binaries, abnormal soft X-ray lines can arise from the different ionization states of the two mini-disks. 
Motivated by the diagnostic capabilities of the reflection spectrum established in single SMBH systems, we now assess the prospects of using X-ray reflection spectroscopy for binary parameter estimation -- a step beyond binary identification.
Given the large number of parameters necessary to characterize binary SMBHs and their accretion flows, any detailed model for the composite binary reflection spectrum robust enough to withstand parametrized fitting would require significant effort to develop.
Before parametrized fitting is developed and attempted, it is therefore prudent to evaluate its ``return on investment" by using a model-independent approach to determine the diagnostic value of the composite X-ray reflection spectrum of an SMBH binary.
With this aim, we investigate the dependence on binary parameters of the most prominent reflection features: the \feka\ profile (rest frame $\sim$ 6.4 keV) and the Compton `hump' (`CH'; peaking around 20 to 30 keV). 
We define and use model-independent spectral metrics to characterize their spectral morphology and search for correlations between the binary parameters and its composite spectrum.

In section \ref{sec:methods}, we outline our methods for computing the binary reflection spectra following the same approach as \citetalias{malewicz2025} and define the spectral metrics used in the following section.
Section \ref{sec:results} presents how binary parameters may be tied to standard spectral metrics for the composite \feka\ profile and the Compton hump from the binary reflection spectrum.
The results are discussed in Section \ref{sec:discussion}, where we
also lay out observational strategies for parameter estimation. Conclusions are found in Section \ref{sec:conclusion}.


\section{Methods}\label{sec:methods}

The composite spectra are computed using the same method as in \citetalias{malewicz2025}, with one notable exception being the added implementation of explicitly spin-dependent radiative efficiencies. 
Using \relxill's lamppost-Comptonization model \texttt{relxilllpCp} \citep{dauser2014,garcia2014}, we create a library of 28,000 synthetic binary spectra by varying individual SMBH spin magnitudes ($a_1$ and $a_2$), binary mass ratio ($q$), total mass accretion rate (\ltot), inclination ($i$) and orbital phase ($f$). We define these parameters in the subsequent section and list their values in Table \ref{tab:param}.

\begin{deluxetable*}{lc|c}
\tabletypesize{}
\tablewidth{0pt} 
\tablecaption{Model parameters. \label{tab:param}}
\tablehead{
\colhead{Parameter} & \colhead{} & \colhead{Values}
} 
\startdata
Mass ratio & $q$ & 0.2,~0.4,~0.6,~0.8,~1.0 \\
Spin parameters & $a_1,~a_2$ & -0.998,~0,~0.5,~0.998  \\
Inclination & $i$ & 15\degree,~30\degree,~45\degree,~60\degree,~75\degree,~ \\
Orbital phase & $f$ & 0\degree,~45\degree,~90\degree,~135\degree,~225\degree,~270\degree,~315\degree \\
Total Eddington luminosity ratio & \ltot & 0.05,~0.15,~0.25,~0.35,~0.45,~0.55,~0.65,~0.75,~0.85,~0.95  \\[5pt]
\enddata
\tablecomments{
A total of $4 \times 4 \times 10 \times 5 \times 7 \times 5 = 28,000$ binary spectra, or 800 physically distinct SMBH binaries (inclination and orbital phase being extrinsic properties of the observer).
We reject any binary where the secondary's accretion rate surpasses the Eddington rate by a factor of 2 or more (97 out of 800, or $\sim 12\%$). 
} 
\end{deluxetable*}

\subsection{Computation of binary spectra}\label{sec:methods-compute}

The binary consists of two SMBHs of masses $M_1$ and $M_2 = qM_1$ at a separation of $s = 100 \rg$ orbiting a common center of mass on concentric circular orbits. Each SMBH is surrounded by its own mini-disk, responsible for the reflection spectrum\footnote{We do not include possible contributions from either the streams or the CBD, as they should be sub-dominant in the hard X-ray regime \citep[$\gtrsim 2$ keV;][]{dascoli2018}.}.
Conversely, the inner truncation radii \rin\ of each mini-disk is set solely by the spin parameter $a\equiv Jc/GM^2$ \citep{kerr1963}.
We assume the each mini-disk is illuminated only by its own corona, suspended 10 \rg\ above the SMBH spin axes\footnote{This assumption for a fixed coronal height is not inconsequential and is discussed in \S~\ref{sec:discussion-limitations}.}.

\begin{figure}
    \centering
    \includegraphics[width=\linewidth]{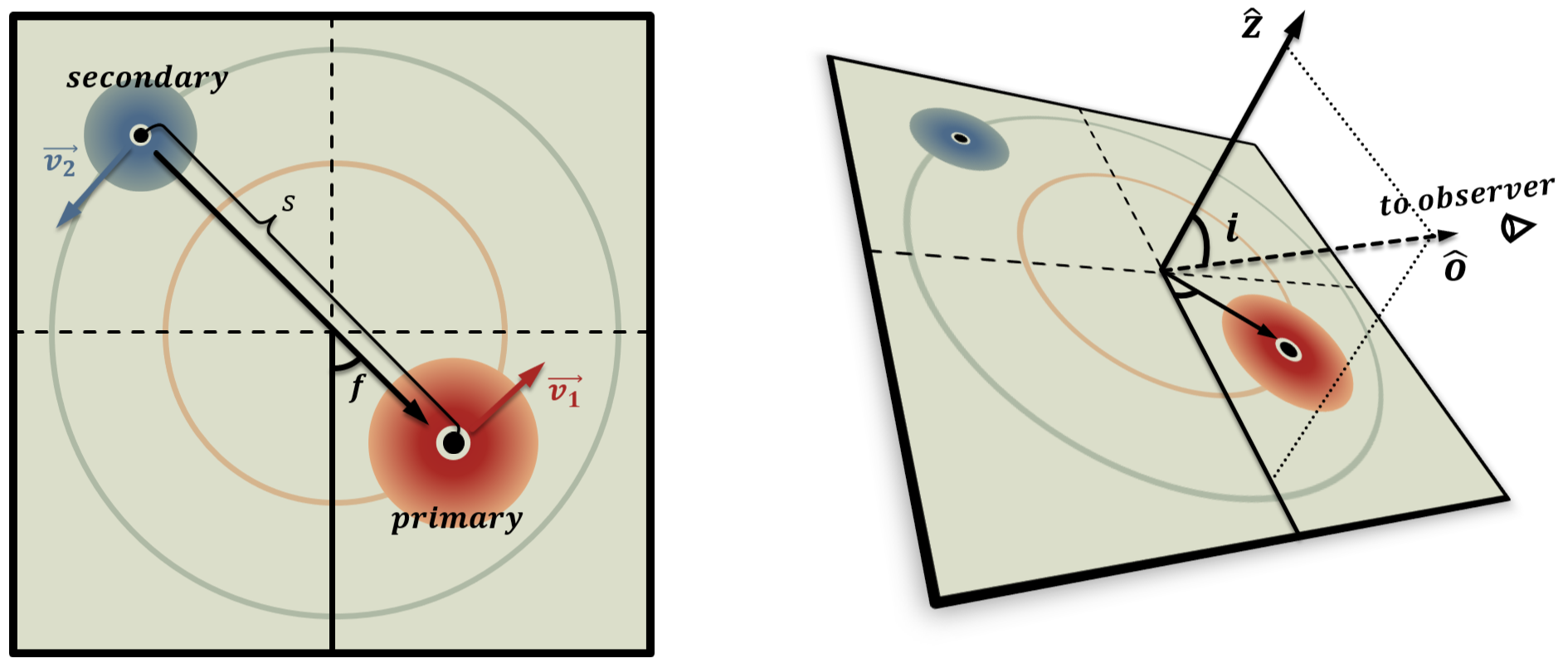}
    \caption{Our SMBH binary model consists of two `mini' accretion disks embedded in the binary orbital plane, characterized by the inclination angle $i$ between the orbital angular momentum vector ${\hat{z}}$ and the line-of-sight vector ${\hat{o}}$. Seen from the tip of the positive $z$-axis, the two black holes are on circular, counter-clockwise orbits described by the phase angle $f$.
    }
    \label{fig:binary_diagram}
\end{figure}

We choose a fully co-planar configuration in which the binary orbital angular momentum and the mini-disk angular momenta are aligned along a unit vector ${\hat{z}}$, which forms an inclination angle $i$ with the observer's line-of-sight ${\hat{o}}$.
The SMBH spin axes are then either aligned ($a>0$) or anti-aligned ($a<0$) with ${\hat{z}}$.
Fig.~\ref{fig:binary_diagram} is a schematic representation of our model geometry. 
%
The two SMBHs orbit their common center-of-mass as a function of the orbital phase $f$, defined as the angle between the projection of the line-of-sight vector ${\hat{o}}$ onto the binary orbital plane and the position of the primary SMBH.

We model the \textit{total} binary accretion rate $\dot{M}_{\rm tot} =\dot{M}_1+\dot{M}_2$ as steady in time, and express it through the Eddington luminosity ratio, \ltot $=L_{\rm tot}/L_{\rm Edd, tot}$.
The assumption that $\dot{M}$ is constant is reasonable for this orbital geometry (see Section 5.3.2. of \citetalias{malewicz2025}, where we justify this assumption).
The less massive secondary SMBH will accrete at a higher rate, as it orbits closer to the inner rim of the CBD, an effect known as accretion inversion, or preferential accretion \citep{artymowicz1994, guenther2002, hayasaki2006, roedig2011, farris2015}. 
We use the fitting formula for $\dot{M}_2/\dot{M}_1$ derived by \citet{kelley2019}. By adding explicitly spin-dependent radiative efficiency parameters, $\eta_1=L_1/\dot{M}_1c^2$ and $\eta_2=L_2/\dot{M}_2c^2$ \citep{novikovthorne1973}, the Eddington ratios of the primary and the secondary can be expressed as:
\begin{equation}
  \label{eq:lambda1}
  \lambda_1 =  {1 + q \over \eta_1 + \kappa  \eta_2} \eta_1  \lambda_{\mathrm{tot}}
\end{equation}
and
\begin{equation}
  \label{eq:lambda2}
  \lambda_2 = {1 +q \over q} {\kappa \over \eta_1 + \kappa  \eta_2 }  \eta_2  
  \lambda_{\mathrm{tot}}.
\end{equation}
where $\kappa(q)=\dot{M}_2/\dot{M}_1$.
We then follow the same approach as \citetalias{malewicz2025} to compute the ionization gradient of each mini-disk due to illumination by a hot, tenuous but compact corona Compton-upscattering thermal radiation from the disk into the hard X-ray. That illumination forms the underlying coronal power-law that is `reflected' by the disk (absorbed, re-emitted and/or scattered) and dominates at higher energies ($E\gtrsim 20$~keV).

We keep all other model parameters and prescriptions equal to those used in \citetalias{malewicz2025}. 
That includes the peak photo-ionization parameter, as per \citet{ballantyne2017}, for a lamppost corona at an effective height of 10 \rg -- which depends on spin and accretion rate.
We include the contribution from the coronal power-law with photon index $\Gamma=2$ in the reflection spectrum. 
The reflection fraction is self-consistently calculated and following \citet{dauser2022}, we allow returning radiation -- i.e., secondary illumination by reflected photons incident onto the disk a second time.
The binary orbit-induced Doppler shift is applied using \xspec's \texttt{zashift}. 

We choose to discard any of the binaries where the secondary exceeds twice the Eddington accretion rate, as that disk would likely be radiatively inefficient \citep[e.g.,][]{abramowic1988} and the reflection model we use is not built for this regime. 
These represent about 12\% of our total sample (3,430 out of 28,000), and are primarily made up of $q=0.2$ binaries.
This is a lenient limit but it allows us to probe the effects of preferential accretion within the margin of error of the \citet{kelley2019} fitting formula.


\subsection{Decomposition into discrete spectral elements}\label{sec:methods-decomp}

\begin{figure*}
    \centering
    \includegraphics[width=0.9\linewidth]{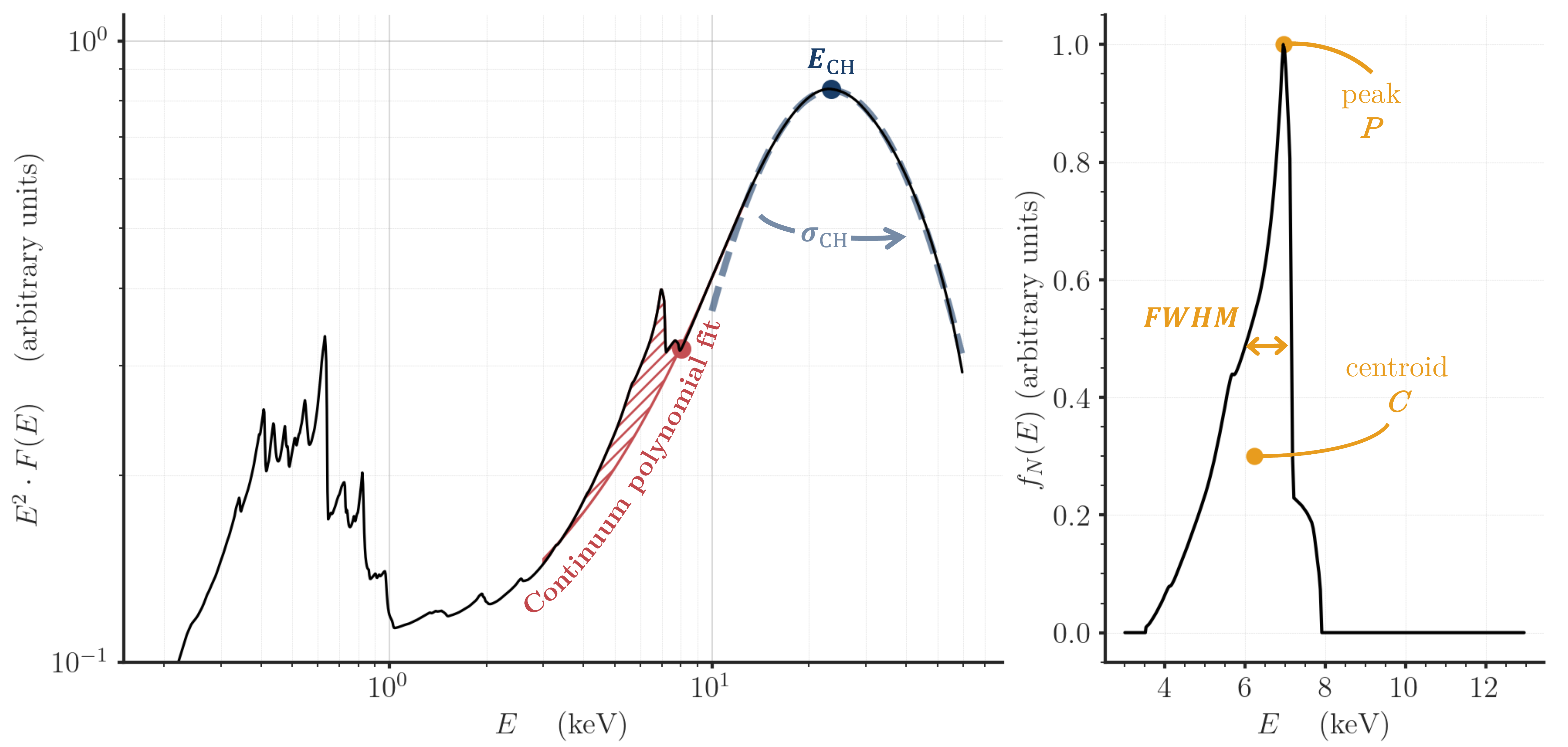}
    \caption{Example composite spectrum for a binary demonstrating the method used to extract the \feka\ profile (3 to 13 kev in the left panel) from the continuum as explained in \S~\ref{sec:methods-decomp}.
    We fit a 4th-degree polynomial (red) to the continuum, which is then subtracted to yield the line profile shown in the right panel. 
    The red point shows the `turnover' energy between the \feka\ profile and the Compton hump.
    In blue, the log-Gaussian fit to the Compton hump. Some of the \feka\ metrics referenced in \S~\ref{sec:methods-metrics} are illustrated in yellow (peak, centroid, and FWHM).
    The source parameters are $\ltot=15\%$, $q=1.0$, $f=0\degree$, $i=45\degree$, $a_1=a_2=0.998$.}
    \label{fig:spectral_decomp_example}
\end{figure*}

In order to quantitatively assess the responsiveness of the composite \feka\ profile and the Compton hump to binary parameters, we devise a scheme to extract them from the spectral continuum, as illustrated in Fig.~\ref{fig:spectral_decomp_example}.
We elect to work with the $E^2 F(E)$ flux (\texttt{eemodel} in \xspec), where $E$ is the photon energy and $F(E)$ the photon flux, which allows reflection features to stand out without factoring out the illuminating coronal power-law.
We define the energy interval of the relativistic \feka\footnote{Depending on the ionization state of the gas, there can be weak contributions from the Fe K$\beta$ and Ni K$\alpha$ fluorescence lines in this energy interval.
Relativistic blurring subsumes the flux from these lines into the larger \feka\ profile. In practice, these weaker satellite lines would not be separable and are therefore included in the metrics calculations.} line as 3 to 13 keV, which is wide enough to fully encompass the broadest profiles in our sample (at high inclination and high spin).
We first identify the `turnover point' between the \feka\ profile and the Compton hump, defined as the highest-energy local minimum to the curve\footnote{Note that in high-$\lambda$ spectra, the disk becomes over-ionized and reflection features are suppressed. For that reason, the turnover point between the \feka\ profile and the Compton hump is no longer a local minimum to the curve. The power-law dominates and needs to be temporarily subtracted in order to identify the turnover point.}. The 2-to-4-keV and turnover-to-18-keV intervals are used to establish a baseline continuum under the \feka\ profile that is fit with a 4th-degree polynomial using \texttt{numpy.polyfit}. These energy ranges provide enough data points for the fit, while avoiding significant contributions from other prominent reflection features (tip of the Compton hump above 18 keV and soft emission lines below 2 keV). Once that continuum is subtracted over the 3 to 13 keV interval, we normalize the \feka\ profile, such that 
the dimensionless flux $f_N(E)$ is defined as:
\begin{equation}
    f_N(E) = \frac{E^2 F(E)}{{P}^2 F({P})},
\end{equation} 
where $P$ is the photon energy at which $f_N(E)$ peaks.


\subsection{Spectral metrics}\label{sec:methods-metrics}

The broad and featureless Compton hump is the simplest spectral element to parametrize. 
It is well-fit with a skewed log-Gaussian of the form:
\begin{equation}
    A  \exp{\left[ -\frac{\ln (E / {E_\textrm{CH}})^2}{2{\sigma_\textrm{CH}}^2} \right]}
\end{equation}
As such, it is characterized by two meaningful quantities: its peak, or median in log-space, ${E_\textrm{CH}}$, and its standard deviation in log-space, ${\sigma_\textrm{CH}}$. 
In qualitative terms, ${\sigma_\textrm{CH}}$ measures how sharply peaked the hump is around ${E_\textrm{CH}}$ and increases when the hump becomes flatter or more skewed.

To characterize the \feka\ profile, we adapt standard, model-independent metrics typically used to describe the shape of a distribution.
Though we considered over a dozen different metrics, we only include in this section the ones that were found to have a tangible diagnostic value.
The \feka\ metrics employed here are used to describe the profile's center, width and asymmetry.

We consider three different metrics to capture the \feka\ center: the centroid ${C}$, the median ${M}$ and the peak ${P}$.
Firstly, the centroid is the flux weighted mean energy in the line interval:
\begin{equation}
    C = \frac{1}{F} \int_i E_i ~ f_N(E_i) ~  d E_i \qquad \text{(keV)}.
\end{equation}
Here, $d E_i=E_{i+1}-E_i = 50$ eV\footnote{For reference, the nominal spectral resolution at 6 keV is $\Delta E\sim150$ eV for XMM-Newton, $\Delta E\sim7$ eV for XRISM's Resolve, $\Delta E\sim2.5$ eV for NewAthena's X-IFU and $\Delta E\sim170$ eV for its WFI.}, and $F$ is the total integrated
flux for the normalized line:
\begin{equation}
    {F} = \int_i f_N(E_i) ~ d E_i \qquad \text{(keV)}.
\end{equation}
A useful metric for characterizing the variability of the composite spectrum is the centroid shift ${\Delta C}= {C}_0 - {C}(f)$, which is the difference between the \feka\ centroid at conjunction ${C}_0$ (when the SMBHs' velocities are orthogonal to the line-of-sight) and the centroid at orbital phase $f$.
The peak energy ${P}$, also defined earlier, is simply the location of maximum $f_N$,
and the median energy ${M}$ cuts the profile in half, such that half of the flux lies on its left and half on its right.

As for the width, we use two measurements: the full width at half maximum, ${FWHM}$, and the variance ${V}$.
The variance ${V}$ is the second moment of the profile's energy distribution:
\begin{equation}\label{eq:V}
    {V} = \frac{1}{{F}} \int_i \left( E_i - {C} \right)^2 ~ f_N(E_i) ~ d E_i \quad \text{(keV}^2).
\end{equation}
We do not consider any higher degree moments, as they are too sensitive to small deviations in the wings of the distribution and would lose much of their diagnostic value when applied to real spectra with noise.

The Pearson asymmetry index,
\begin{equation}
    {AIP} = \frac{{C}-{M}}{\sqrt{{V}}} \quad \text{(dimensionless)}\label{eq:AIP},
\end{equation}
is used to diagnose asymmetry in the bulk of the profile by comparing its centroid to its median. A greater (more positive) value for $AIP$ typically signals a distribution skewed `blue'. The relativistic \feka\ line is already skewed and broadened by relativistic effects in single SMBH systems, which gives it its diagnostic value in the first place. This complicates the mapping between simple skewness indices and the line's actual morphology. As such, the absolute value of this index is less informative than the difference between the mean indices of two different populations.
%


\section{Results}\label{sec:results}

\begin{figure*}[h!]
    \centering
    \includegraphics[width=1\linewidth]{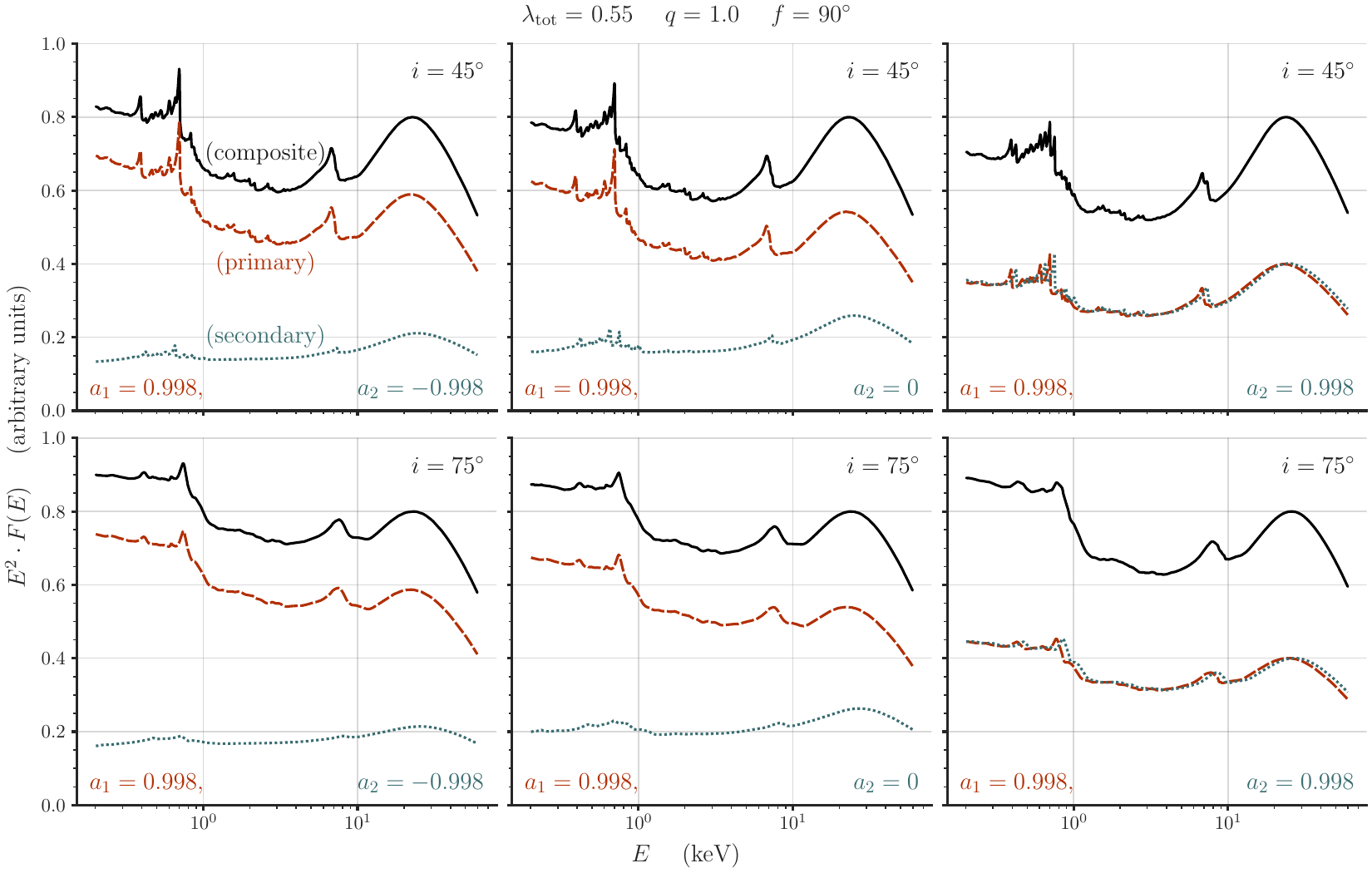}
    \caption{Examples of composite spectra obtained by adding the spectra of the primary (red, dashed) and secondary SMBH (blue, dotted) of an equal-mass binary, 
    varying inclination (top row: $i=45\degree$; bottom row: $i=75\degree$) and spin of the secondary SMBH (from left to right: $a_2=-0.998,~0,~0.998$). 
    The total luminosity output of the binary is fixed at 55\% of the Eddington rate,
    and the orbital phase is $f=90\degree$ (opposition, secondary spectrum blue-shifted). 
    The relativistically broadened \feka\ profile is visible in all spectra around 7 keV, and the Compton hump describes the broad `bump' to the right, peaking around 25 keV.
    Note that each panel's y-axis was rescaled to have the composite's Compton hump peak to 0.8 for visual clarity.}
    \label{fig:spectral_addition_example}
\end{figure*}

\subsection{Binary spectra}\label{sec:results-spectra}

We explore in \citetalias{malewicz2025} the effects of mass ratio and total mass accretion rate on the composite reflection spectrum. Two important parameters that were not investigated then are the observer inclination angle, $i$, and the two SMBH spins, $a_1$ and $a_2$.

Figure \ref{fig:spectral_addition_example} shows the spectra from six different spin and inclinations configurations of an equal mass ($q=1.0$) binary. 
Each row corresponds to a different binary plane inclinations ($i=45\degree$ at the top and $75\degree$ at the bottom.) The two SMBHs are at opposition ($f=90\degree$), where the primary's emission is redshifted.
The primary SMBH (red, dashed) has its spin fixed at $a_1=0.998$, but we vary the secondary SMBH's (blue, dotted) spin $a_2=-0.998,~0,~0.998$ between the three columns. 
A lower spin is associated with a larger ISCO (innermost stable circular orbit) radius, truncating the innermost disk regions responsible for the bulk of the reflection flux. This results in a significant drop in luminosity compared to an identical but higher-spin BH. In our model, this is expressed through the radiative efficiency $\eta$.
The difference in luminosity between the primary and the secondary is starkest in the case where $a_1=0.998$ and $a_2=-0.998$ (top-left panel), due to the factor of almost 10 between their radiative efficiencies ($\eta_1=0.321$ but $\eta_2=0.038$), despite a mass ratio $q=1$.
The innermost disk regions also tend to be more ionized and, given their proximity to the SMBH, more gravitationally redshifted. This affects both continuum emission and spectral features like the \feka\ profile. We discuss spin effects on the composite spectrum in more detail in Sections \ref{sec:effective-spin} and \ref{sec:individual-spin}.

The inclination angle is an extrinsic property of the SMBH binary system, but due to the extremely relativistic spacetime around the SMBH, observer effects can drastically alter how the reflection spectrum appears.
A higher inclination increases the line-of-sight velocities of the accretion flow, and thus the magnitude of the Doppler effect on the emergent spectrum. This contributes to an apparent widening of otherwise narrow spectral features.


\subsection{Inclination angle of the orbital plane, $i$}\label{sec:inclination}

Measuring the relativistic \feka's peak $P$ is straightforward, if it can be isolated from the narrower components to the profile originating at larger radii (see \S~\ref{sec:discussion-limitations}).
\begin{figure}
    \centering
    \includegraphics[width=0.9\linewidth]{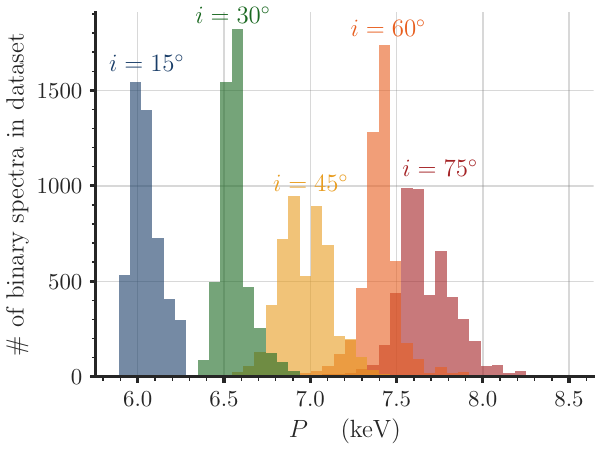}
    \caption{
    Distribution of the \feka\ peak energies for spectra grouped by disk inclination in our synthetic SMBH binary dataset ($i \in [15\degree,~30\degree,~45\degree,~60\degree,~75\degree]$). 
    The peak ${P}$ of the composite \feka\ line shifts systematically to higher energies with increasing inclinations, with typical peak energies of $\sim$6.1, 6.6, 7.0, 7.4 and 7.6 keV, respectively. Each distribution exhibits a characteristic spread $\sim$0.5 keV.
    }
    \label{fig:inclination}
\end{figure}
%
Figure \ref{fig:inclination} shows a histogram of values for the \feka\ peak ${P}$ in our full binary dataset, separated by inclination angles ($i=15\degree,~30\degree,~45\degree,~60\degree$ and $75\degree$). 
%
We find that, if the binary mini-disks are coplanar with the binary orbit, we recover the well-known relationship between the relativistic \feka\ peak (coincident with its blue wing's drop-off) and the disk's inclination with respect to the line-of-sight. A more edge-on viewing angle enhances the magnitude of the Doppler shift acting on the approaching lobe of the (mini-)disk, boosting and blueshifting the fluorescent emission away from its usual rest-frame energy of $\sim 6.4$ keV into the brighter, bluer wing of the double-horned line profile \citep[e.g.,][]{fabian1989}.

Overall, the inclination is arguably the single most influential parameter in shaping the observed relativistic \feka\ profile, and the dependability of the line peak in measuring it is well documented for single SMBHs
\citep[e.g.,][]{cunningham1975,pariev2001,reynolds2003, brenneman2006, gates2020}. 
Figure \ref{fig:inclination} illustrates that this diagnostic remains applicable 
to binaries and appears largely insensitive to parameters like mass ratio and spin. Moreover,
at orbital separations of 100~\rg, the Doppler shift induced by the binary's orbital motion is small ($\la 0.3$ keV). 
The diagnostic value of the peak measurement therefore does not appreciably deteriorate when varying
orbital phase, nor does it weaken when varying the accretion rates\footnote{In highly ionized disks, \feka\ fluorescence can become dominated by H- and He-like Fe ions with a rest frame K$\alpha$ energy $\sim 6.70-6.97$ keV \citep[e.g.,][]{kallman2004, garcia2010}. Even though we consider a wide range of ionization fractions, this effect does not supersede the diagnostic value of the composite \feka\ peak for measuring inclination. For example, the slight bimodality of the $i=45\degree$ distribution is due in part to ionization effects.}.
Prospects for constraining binary inclination from a single epoch spectrum are promising, which may lay the foundations for more precise measurements of other parameters, like mass ratio or spin.


\subsection{Binary mass ratio $q$}\label{sec:mass-ratio}

\begin{figure*}[t!]
    \centering
    \includegraphics[width=\linewidth]{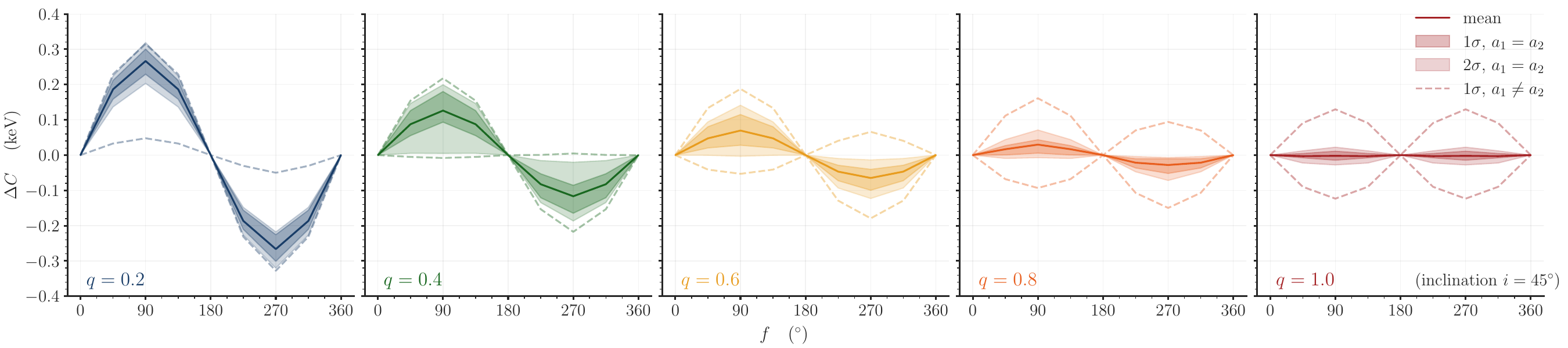}
    \caption{Average motion of the \feka\ centroid of a binary SMBH spectrum throughout its orbit for $i=45\degree$. Each panel corresponds to a given mass ratio (from left to right, $q=0.2,~0.4,~0.6,~0.8,~1.0$). Also included are the $1\sigma$ and $2\sigma$ intervals encompassing 68\% and 95\% of the binaries where $a_1=a_2$ (solid lines) and the $1\sigma$ interval for binaries where $a_1 \neq a_2$ (dashed).
    In unequal mass binaries, the brighter secondary easily dominates the behavior of the composite \feka\ profile centroid (blueshifted at $f=90\degree$, when the secondary is approaching). For $q=0.2$, the average magnitude of the blueshift peaks to a value of $\sim0.3$ keV at $f=90\degree$.
    }
    \label{fig:mass_ratio_5_panels}
\end{figure*}

If information about the orbital inclination can be obtained from the binary spectrum, it could in principle be used as a prior that would inform estimation of other parameters. In this section, we assume that inclination is known ($i=45\degree$) for simplicity, and consider the correlation between the binary mass ratio and location of the \feka\ line profile. Figure \ref{fig:mass_ratio_5_panels} illustrates one time-variable aspect of the binary composite spectrum: the Doppler shifting of the \feka\ profile centroid. 
Each panel corresponds to the average \textit{centroid shift} ${\Delta C}$ for binaries with a given mass ratio. ${\Delta C}$ tracks the bulk motion of the composite \feka\ profile through a full orbit, and could in principle be measured using multi-epoch spectra over the orbital timescale of the binary.
The motion of the composite profile is largely driven by that of the more luminous of the two SMBHs and can be used to determine the orbital phase of the binary.
If spins $a_1\simeq a_2$, as the binary mass ratio decreases, the less massive secondary, accreting at a higher rate, increasingly dominates the variability of the binary \feka. This is evident in the average sinusoidal motion of the centroid ${\Delta C}$ throughout the orbit matching the motion of the secondary.

A lower mass ratio also implies a greater orbital velocity of the secondary SMBH (at $s=100~\rg$, up to a few percents of $c$, the speed of light), enhancing Doppler shifting and boosting. 
The amplitude of that shift is inversely correlated with $q$. It is also directly proportional to $\sin{i}$ and will increase as the orbital separation between the SMBHs shrinks. The centroid shift for an equal mass binary predictably cancels out to 0 due to the equal and opposite contributions of both SMBHs. In such cases, the width of the \feka\ profile may increase, but the two constituent profiles may not always shift enough to observe a clear double peak.

There are other factors that influence the brightness of the two mini-disks relative to one another, however.
An example of this is the spin dependence of radiative efficiencies. This effect is illustrated in Fig.~\ref{fig:mass_ratio_5_panels}: the statistical spread around the average motion of the centroid is increased when considering the full spin sample, which includes binaries where $a_1=0.998$ and $a_2=-0.998$. The left-most panels of Figure \ref{fig:spectral_addition_example} show that in those extreme cases, the bolometric luminosity of the secondary mini-disk is greatly suppressed.
Similarly, in unequal mass ratios, the more rapidly accreting SMBH may approach the radiatively inefficient super-Eddington regime ($\lambda_2 \gtrsim 1$), where a decreased fraction of its total bolometric luminosity is emitted in the X-rays.
For that reason, including binaries where $\lambda_2>2$ in our sample also lowers the magnitude of the average centroid shift for low mass ratios.
Still, $\Delta C$ shows minimal dependence on total accretion rate or SMBH spin magnitudes (as long as $a_1 \simeq a_2$).


\subsection{Effective black hole spin, \chieff}\label{sec:effective-spin}

X-ray reflection spectroscopy is uniquely promising for constraining spin, as it probes emission down to the innermost stable circular orbit (ISCO). This section investigates if imprints of the individual SMBH spins remain in the composite binary spectrum, and thus whether they might be recoverable through broadband parametrized fitting.

Because the parameter space is bound to be very degenerate, we consider the impact of a simple scalar analog to the binary average spin commonly used in GW measurements, the effective spin \chieff
.
The effective spin of a binary is commonly defined as the mass-weighted combination of the individual spins projected along the binary orbital momentum:
\begin{equation}
    \chieff = \frac{\chi_{1z} + q \, \chi_{2z}}{1+q}.
\end{equation}
In a co-planar geometry, where both mini disks are aligned with the individual BHs' angular momentum and the binary orbital momentum, $\chi_{iz}\equiv a_i$. 
\added{We note that throughout this section, we use `low' to refer to negative (or retrograde) values of either $\chieff$ or $a_i$ (with an inner disk truncation radius $\risco\approx9\rg$), as opposed to `high' spin, which is always prograde with $\risco\approx1.23\rg$.}

\subsubsection{Compton hump}\label{sec:effective-spin-CH}

Though the \feka\ has historically been the most obvious spectral feature used in spin measurements, standard practice is now to fit broadband spectra to break intrinsic model degeneracies \citep[e.g.,][]{dauser2014,bambi2021, garcia2026}.
The Compton hump, for example, is also shaped by the innermost disk regions, which may become highly photo-ionized due to illumination by a central corona. The radial extent of these disk regions is set by the inner disk truncation radius, which in most reflection models is often assumed to be solely a function of BH spin. Therefore, the hump's shape may be providing supplementary information about the SMBH spin parameter and help resolve intrinsic model degeneracies.
To investigate this hypothesis, we examine the responsiveness of the log-Gaussian fit to the hump to changes in the effective spin of the binary.

\begin{figure*}
    \centering
    \includegraphics[width=\linewidth]{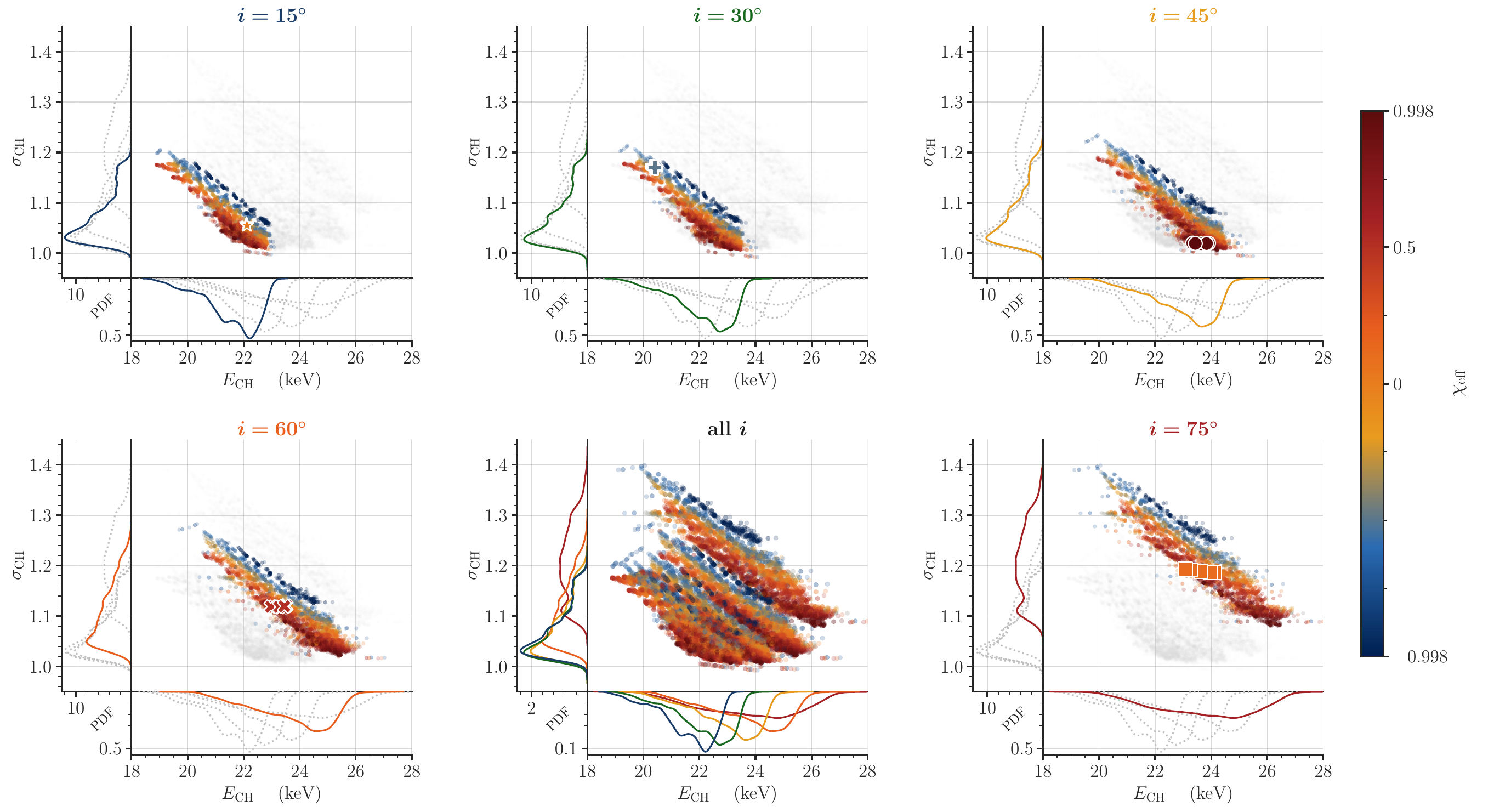}
    \caption{
    Compton hump peak (${E_\textrm{CH}}$) vs. log-width values (${\sigma_\textrm{CH}}$) for all binary spectra in our sample, color-coded by effective spin $\chi_{\rm eff}$, and separated by inclination angle (from top-left to bottom-right: $i=15\degree,~30\degree,~45\degree,~60\degree$ and $75\degree$ with all $i$ combined in the center bottom panel). 
    The full distribution is shown in grey in the background of each scatter plot for reference.
    Marginal Probability Density Functions (PDF) projected onto the horizontal (${E_\textrm{CH}}$) and vertical (${\sigma_\textrm{CH}}$) axes illustrate the 1D distribution for each inclination sub-population.
    Finally, special symbols show the expected orbit-induced Doppler shift variation in CH morphology for one representative binary in each inclination bin, with parameters listed in Tab.~\ref{tab:chparam}.
    }    
    \label{fig:compton_spin}
\end{figure*}

We find that $E_\text{CH}$ and $\sigma_\text{CH}$ show sensitivity to the binary's effective spin \chieff,
suggesting that a broadband parametrized fitting scheme that captures the hump's peak may be able to extract information about the spin parameters of a SMBH binary.
Figure \ref{fig:compton_spin} shows the distribution of Compton hump peak (${E_\textrm{CH}}$) and log-width values (${\sigma_\textrm{CH}}$) for all binary spectra in our sample. Each point is colored according to the value of \chieff. It illustrates that a lower \chieff\ is typically associated with larger ${\sigma_\textrm{CH}}$ and lower ${E_\textrm{CH}}$.
In this case, because the spectral metric is sensitive to asymmetry by construction, larger ${\sigma_\textrm{CH}}$ values are indicative of a more skewed hump. Notably, this trend persists across orbital phases and inclination.

The Compton hump is not an emission feature in the same way the \feka\ line is. While the \feka\ line's shape is attributable in large part to relativistic kinematics, the morphology of the CH reflects the complex physical conditions of the accretion disk gas itself. Specifically, the hump is bounded on the low-energy end by the photoelectric absorption edges associated with iron \citep[e.g.,][]{kallman2004} and nickel K$\alpha$ and K$\beta$ transitions -- the exact thresholds of which are very sensitive to ionization levels. Conversely, the high-energy side of the hump is made up of photons having undergone Compton scattering with free disk electrons, which produces a small excess of photons in the 20 to 30 keV range. This gives this part of the spectrum the appearance of a `hump' \citep{lightman1988}.

The high-energy side of the hump, spanning over 30 keV in range, is therefore largely insensitive to spin. Its low-energy edge, however, is critically dependent on the ionization state of the reflecting disk material.
In centrally concentrated coronal geometries, the disk ionization follows a steep radial gradient, with the innermost disk receiving the brunt of the hard coronal irradiation \citep[e.g.,][]{shakurasunyaev1973,fukumura2007,ballantyne2017}. These regions are far more likely to Compton-reflect coronal photons than to absorb them. A lower spin results in a truncation of those highly reflective regions from the disk, leaving only the less ionized material at larger radii that is responsible for a deeper photoelectric absorption gap.
A possible interpretation for the behavior observed in Fig.~\ref{fig:compton_spin} would be that 
high spins conserve most of the bright, highly ionized inner disk, which
fills in the photoelectric absorption gap, restoring some symmetry to the hump, steepening the peak and returning a `narrower' fit.
In the binary case, the effective spin is simply the mass-weighted average of the two SMBH spins and is therefore dominated by the more massive primary. Because the primary typically accretes at a lower rate, its mini-disk is less likely to be highly ionized. As a result, whether the primary's innermost disk regions are truncated or not can visibly alter the depth of the photoelectric absorption gap in the composite spectrum. Indeed, replacing the effective spin with the individual spins ($a_1$ or $a_2$) in Fig.~\ref{fig:compton_spin} reveals a stronger dependence on $a_1$ than on $a_2$. Similarly, we find that when mapping each constituent spectrum (primary and secondary) onto the ${E_\textrm{CH}}$-${\sigma_\textrm{CH}}$ parameter space, the primary shows a clearer separation between spin values that the secondary.

In Figure \ref{fig:compton_spin}, we also included an example binary for each inclination, for which we plot ${E_\textrm{CH}}$-${\sigma_\textrm{CH}}$ at every orbital phase (see symbols in legend and in Table \ref{tab:chparam}). This allows us to assess how the appearance of the Compton hump may change due to orbital Doppler shifting for different binary parameters. At low inclinations ($i=15\degree$, represented by {\color{star}  $\Star$}), the orbital effects are imperceptible, owing primarily to the low line-of-sight velocities of each mini-disk. At $i=30\degree$ ({\color{plus}  $\Plus$}), the $q=1.0$ binary emission is dominated by the secondary, due to its higher spin. The peak of the Compton hump shifts by about 0.5 keV and the difference in its log-width between $f=90\degree$ and $f=180\degree$ is of about $5\%$. At $i=75\degree$ ({\color{square} $\blacksquare$}), though the mass ratio is 0.8, the two SMBH spins are maximally prograde and retrograde, respectively. This means that the primary easily dominates the emission, and the composite CH peak shifts by $\sim 1.3$ keV throughout the orbit, even if the log-width evolution is minimal.

\begin{deluxetable}{c||cc|cc|c}[h]
\tabletypesize{}
\tablewidth{0pt} 
\tablecaption{Source parameters for each cluster of points in  Figs.~\ref{fig:compton_spin} and ~\ref{fig:feka_centroid_vs_var} \label{tab:chparam} (the color of each symbol is set by the \chieff\ value).}
\tablehead{
\colhead{$i$} & \colhead{\ltot} & \colhead{$q$} & \colhead{$a_1$} & \colhead{$a_2$} & \colhead{$\chi_{\rm eff}$}
} 
\startdata
$15\degree$ ({\color{star}  $\Star$})  & $45\%$ & $0.6$ &  $0$ & $0$ & $0$ \\
$30\degree$ ({\color{plus}  $\Plus$}) & $95\%$ & $1.0$ &  $-0.998$ & $0$ & $-0.5$ \\
$45\degree$ ({\color{circle} $\Circle$}) & $15\%$ & $0.4$ &  $0.998$ & $0.998$ & $0.998$ \\
$60\degree$ ({\color{cross} $\Cross$}) & $25\%$ & $0.2$ &  $0.5$ & $0.5$ & $0.5$ \\
$75\degree$ ({\color{square} $\blacksquare$}) & $55\%$ & $0.8$ &  $0.998$ & $-0.998$ & $0.111$ \\[5pt]
\enddata
\end{deluxetable}

In summary, we find a correlation between the shape of the Compton hump and the binary effective spin, which is likely mediated by the depth of the photo-electric absorption gap to the left of the Compton hump peak. Higher values of \chieff, which increase the contribution of the innermost mini-disks to the reflection spectrum, are associated with lower $\sigma_{\rm CH}$ and \added{higher} $E_{\rm CH}$, as illustrated in Figure \ref{fig:compton_spin}.
This underlines that spin effects are broadband, and that there is definitive value in obtaining as complete of a reflection spectrum as possible for parameter estimation, especially when trying to constrain spin. The Compton hump alone is not sufficient to estimate spin, but it may prove useful for breaking model degeneracy.

\subsubsection{\feka\ profile}\label{sec:effective-spin-feka}

\begin{figure*}
    \centering
    \includegraphics[width=\linewidth]{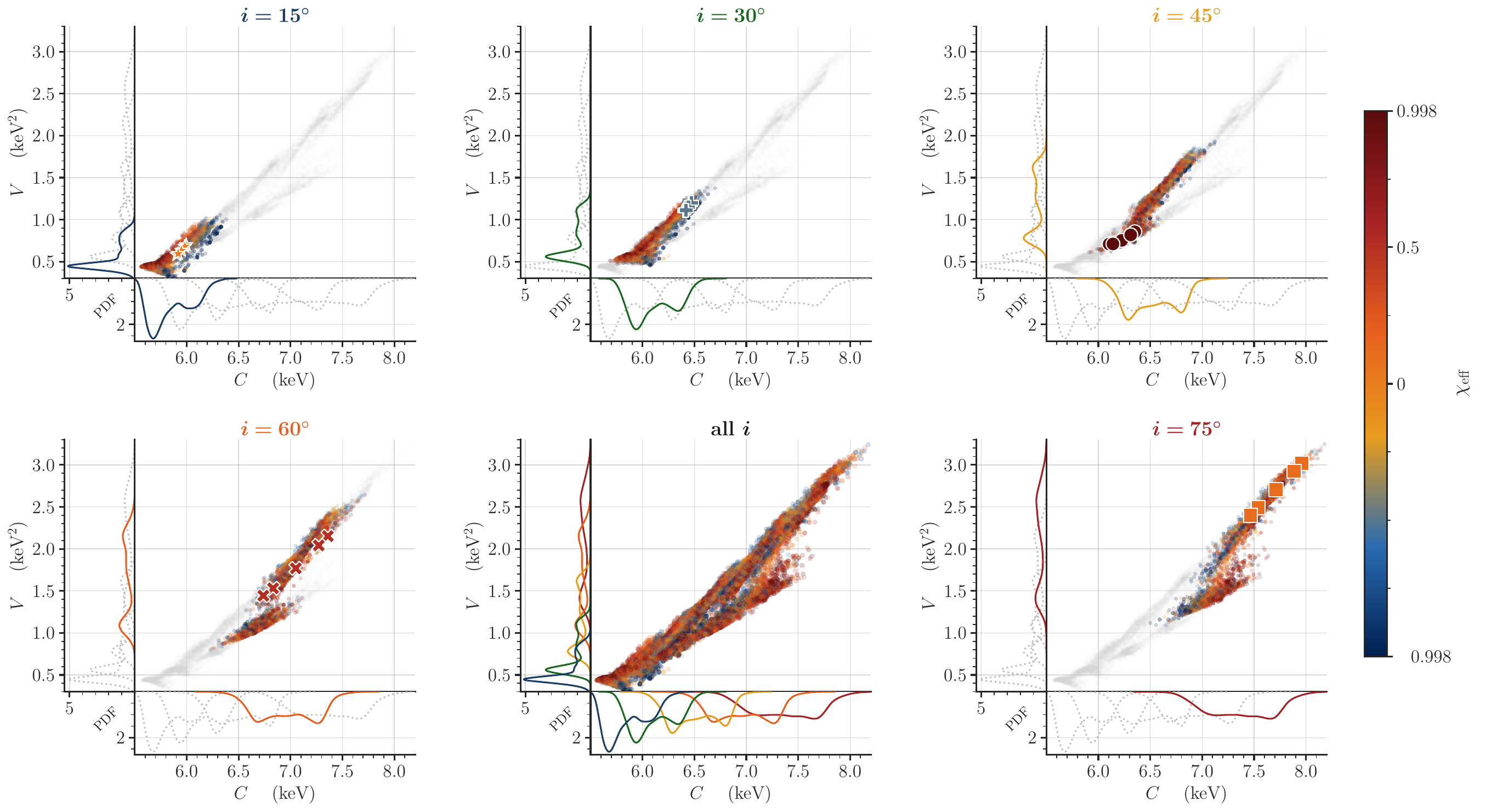}
    \caption{As in Fig.~\ref{fig:compton_spin} but for \feka\ variance ${V}$ vs. centroid ${C}$. Like in Fig. \ref{fig:compton_spin}, each cluster of a given colored symbol corresponds to a single example binary over its full orbit (one for each inclination angle, see Tab.~\ref{tab:chparam} for their binary properties). 
    }
    \label{fig:feka_centroid_vs_var}
\end{figure*}

We are also able to reproduce known spin diagnostic tests of the relativistic \feka\ profile, but with the composite profile and binary effective spin instead.
Indeed, in single BHs, the imprint of spin onto the \feka\ profile is dependent on the viewing angle of the thin accretion disk.
For all moderate to low inclinations, the diagnostic criterion for spin is the extent of the profile's red wing, which is a proxy for the radius of greatest gravitational redshift (i.e., the innermost stable circular orbit, ISCO).
Conversely, for high inclination angles (more edge-on), the Doppler boosting of the approaching half disk becomes significant enough as to turn the blue wing into the dominant determinant of the line centroid. Under a Keplerian velocity profile for the accretion disk, the magnitude of that Doppler shift/boost increases with proximity to the central BH -- therefore turning the blue wing's drop-off into an effective way to estimate BH spin.

That same behavior is illustrated in Fig.~\ref{fig:feka_centroid_vs_var}, which shows the location on the \feka\ ${C}$-${V}$ plane of all binaries in our sample, colored as a function of their effective spins.
The two inclination regimes are displayed clearly in the $i=15\degree/30\degree$ (top left and middle; narrower with a redder centroid) and the $i=75\degree$ panel (bottom right; broader with a bluer centroid). The concentration of points along the $x=y$ diagonal points to the fact that centroid and variance are highly correlated properties. This is due in large part to the asymmetry of the \feka\ profile, whose dominant blue wing will shift the centroid bluer as the profile widens.
However, in both the $i=15\degree$ and $i=\added{3}0\degree$, the dependence on effective spin ($\chieff$) is orthogonal to this trend. As spin increases in general, profile broadening is driven almost entirely by the extension of the gravitationally redshifted red wing. As a consequence, higher effective spins produce larger \feka\ variances $V$ paired with redder centroids $C$, causing high-\chieff\ binaries to populate the wide, red edge of the $C-V$ distribution.
The opposite tendency is exhibited by the $i=75\degree$ binaries:
the red edge of the distribution is preferentially populated by \textit{low}-\chieff\ points, with the high-\chieff\ ones favoring the blue-most edge.
This is consistent with the enhanced blue wing of the \feka\ profile expected from rapidly spinning SMBHs observed at high inclinations.

The bottom middle panel, which shows the placement of all the binaries in the sample (i.e., all inclinations), shows a high degree of degeneracy, and helps illustrate why spin estimation requires broadband parametrized fitting.
We can also observe the Doppler variations in individual binaries over their orbit, wherein a single binary may travel along the diagonal of the distribution -- orthogonal to the effective spin dependence. This is due to the fact that the variance is defined with respect to the centroid (equation \ref{eq:V}), which does shift during the binary orbit, as shown in Figure~\ref{fig:mass_ratio_5_panels}.


\subsection{Individual SMBH spins, $a_1$ and $a_2$}\label{sec:individual-spin}

\begin{figure}
    \centering
    \includegraphics[width=\linewidth, trim={0 6.5cm 0 0}, clip]{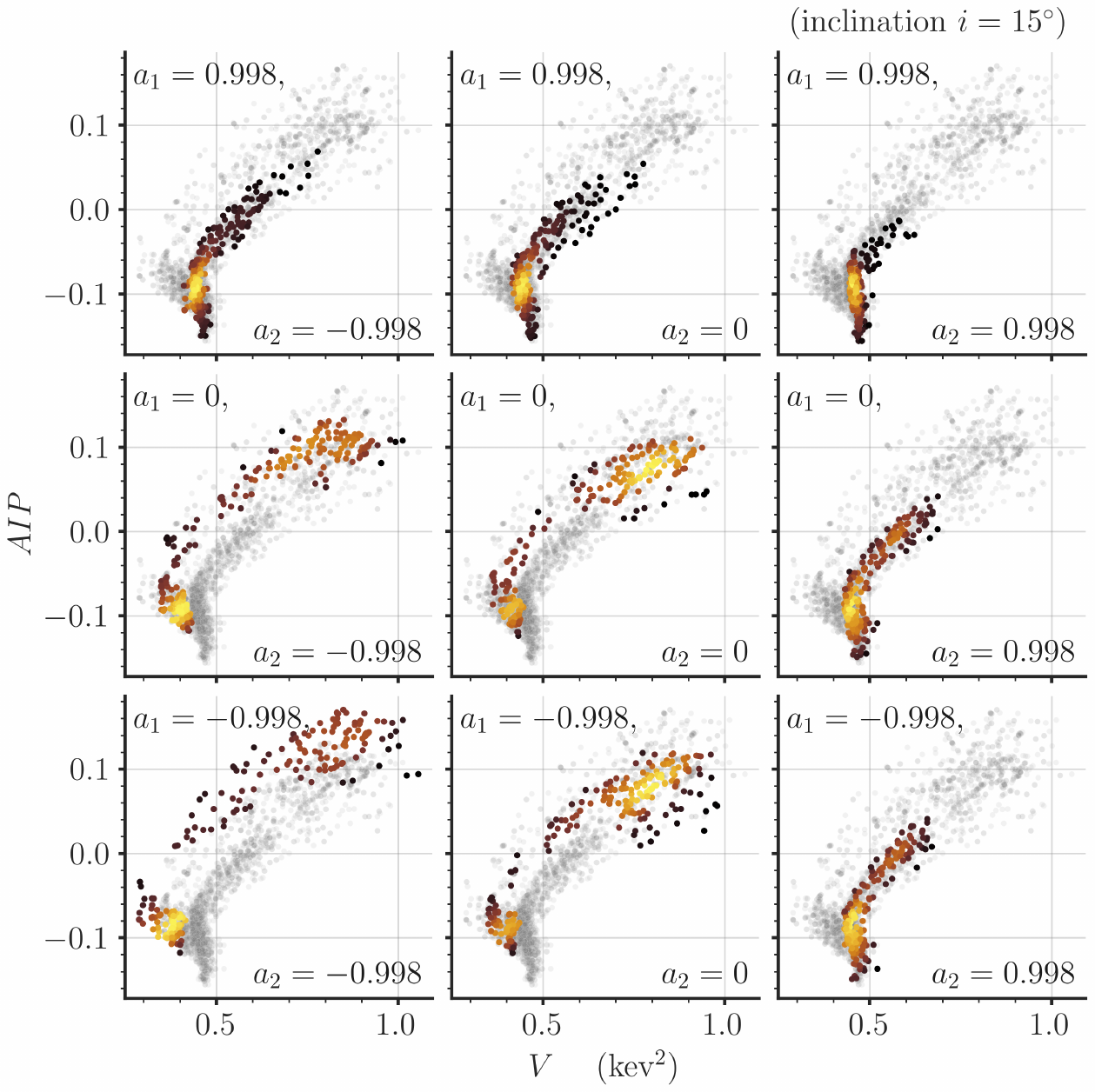}
    \caption{We vary individual SMBH spins ($a_{1,2} \in [-0.998,0,+0.998]$) and map their profile in the width vs. asymmetry plane (using variance ${V}$ for width and Pearson index ${AIP}$ for asymmetry). Inclination is fixed at 15\degree. The grey dots show the full spin sample for comparison and are identical in all 9 panels. The color gradient (from dark red to bright yellow) is a qualitative visualization of the distribution density computed via Gaussian kernel density estimates, where the peaks of the distributions are in yellow. 
    }
    \label{fig:feka_spin_grid}
\end{figure}

In Figure \ref{fig:feka_spin_grid}, we look at the impact of individual SMBH spins on the shape of the \feka\ profile -- namely, variance ${V}$ on the $x$-axis and asymmetry index ${AIP}$ on the $y$-axis.
Each individual plot corresponds to a specific ($a_1,a_2$) subset of the larger $i=15\degree$ population. Rows correspond to different values of $a_1$, and columns do the same for $a_2$.
Grey dots represent the full $i=15\degree$ sample for comparison, and the colors highlight the loci of high population density (in yellow), computed individually for each plot using Gaussian kernel density estimates with \texttt{pandas.kde}.

This highlights that while a fair amount of degeneracy is still expected in the simplest of cases (where both disks are coplanar with the BH spins and binary orbit, and at a low inclination, $i=15\degree$, where the \feka\ line profile appears sharper), the composite spectrum of a binary containing a high-spin SMBH should still carry the distinctive markers of \feka\ emission close to the ISCO.
Indeed, each of the 9 panels belong to one of 2 distinct regimes where the distribution of points is identical -- the four panels in the lower left corner (where $a_1$ and $a_2=-0.998$ or $0$) are associated with a higher asymmetry index compared to the distribution in the outer, upper-right panels (where either $a_1$, $a_2$ or both $=0.998$). There are two distinct clusters in the moderate-to-low spin case: one with negative ${AIP}$ and ${V}\simeq0.35$ keV$^2$ (narrower than in the high-spin case) and a second one with positive ${AIP}$ and ${V}\simeq 0.6$ to 1.0 keV$^2$.
Referring back to the definition for the ${AIP}$ in Eq.~\ref{eq:AIP}, ${AIP}<0$ means that the centroid is less than the median, which can indicate the presence of an extended red tail, biasing the centroid to the left.
Conversely, without the strong gravitational redshifting typically associated with high spin, disk geometry by default should yield a positive AIP, because relativistic Doppler boosting enhances the blue wing from the approaching half disk.

This is analogous to the single SMBH case, where what can be inferred from the \feka\ profile is not a spin measurement \textit{ per se}, but instead an upper limit on the inner truncation radius necessary to explain the degree of relativistic broadening observed. In the binary regime, this simple variance vs. asymmetry measurement is likely picking up on the presence (or absence) of a significant red tail in the composite profile, which can only be explained by strong gravitational redshifting. 
It may be therefore be possible, in some select cases, to distinguish composite \feka\ profiles from binaries where at least one of the component SMBH is spinning near the maximal rate, from the ones where both spins are moderate to low (or retrograde) -- but that it will likely be difficult to place more precise constraints.

Spin imprints on the \feka\ profile are already subtle for single SMBHs and become further diluted in a composite binary line profile.
Still, in some cases like the one in Fig.~\ref{fig:feka_spin_grid}, we find that the metrics we use to describe the composite line profile show measurable sensitivity to its constituent SMBH spins, suggesting that broad spin constraints may be achievable. Such constraints will require broadband coverage (including the Compton hump peak, as per \S~\ref{sec:effective-spin}) to help break the degeneracies inherent to spin measurement in single SMBHs and the ones brought about by the addition of a secondary SMBH.


\section{Discussion}\label{sec:discussion}

\subsection{Possible strategies for parameter estimation}

The main finding of this work, that different spectral metrics correlate preferentially with different binary parameters, is encouraging and merits further investigation.
This indicates that the investigated SMBH binary parameters are not fully degenerate with one another and could to some degree be isolated and extracted iteratively. 
If this phenomenology is found to hold more broadly in the parameter space of SMBH binaries, one can envision that preliminary parameter estimation can be performed in multiple steps as follows:

\paragraph{Inclination}
    The peak of the \feka\ profile in a single-epoch spectrum provides an estimate of the inclination of the mini-disks with respect to the observer's line-of-sight, as illustrated in Figure \ref{fig:inclination}.
    Any additional epoch of spectra may be used to narrow down uncertainties in the initial measurement, but should not be necessary for a first estimate.
    An approximate inclination measurement is a prerequisite to any subsequent diagnostic, as it factors into the Doppler shift on all disk emissions, whether due to binary orbital motion or gas rotation in the mini-disks themselves.

\paragraph{Mass ratio, orbital phase}
    With inclination constrained, that first spectrum can also initiate a time series of \feka\ centroid shift measurements. 
    Figure \ref{fig:mass_ratio_5_panels} shows that the binary \feka\ centroid shifts with the orbital phase, with a magnitude inversely correlated with mass ratio.
    The usefulness of that diagnostic is contingent on obtaining multiple epochs of spectra \added{across distinct orbital phases}, 
    as too few data points would confuse phase with magnitude.
    If concurrent GW measurements are able to identify the orbital phase and/or frequency, the centroid shift in the composite \feka\ can help constrain the likely mass ratio of the source. 
    Alternatively, if estimates for the orbital frequency and mass ratio are obtained from `chirping' in the GW signal (the rapid rise in GW frequency associated with the late inspiral and merger, as might be for the case for some LISA binaries, see \S~\ref{sec:discussion-multimessenger}), X-ray spectroscopy can be used to improve precision, and to test semi-analytic predictions for $\dot{M}_2/\dot{M}_1$ (see \S~\ref{sec:discussion-limitations}).
   If no shift is observed, one may still entertain the possibility of a binary with $q=1.0$. In this case, the width of the \feka\ profile may still vary every binary half-period, as the equal and opposite Doppler shifts pull on the two constituent \feka\ lines.
   Alternatively, if $q\lesssim 0.1$ and the composite \feka\ is dominated by the slow moving primary, the centroid shift may not have sufficient diagnostic power to constrain the orbital phase or the mass ratio.
\paragraph{Spins}
    Spin effects are broadband, and will surface most prominently in the \feka\ profile as relativistic deformation (as hinted at in Figure \ref{fig:feka_centroid_vs_var}), but also in other parts of the composite spectrum (such as the Compton hump, see Figure \ref{fig:compton_spin}). 
    The spectral metrics approach like the one in this work may be inadequate for precise spin measurements, but demonstrates that the composite reflection does display some sensitivity to the spins of its constituent SMBHs. It also shows that while there is significant degeneracy to contend with (as in Figure \ref{fig:feka_spin_grid}), one may still be able to qualitatively assess whether the composite carries the imprint of high spin. The increased radiative efficiency associated with high spin will make it easier for those effects to surface in the composite.
    This supports the idea that information about the binary's spin configuration may be accessible via broadband parametrized fitting.
    Multi-epoch spectra may help to disentangle the contributions from either SMBH, given that their two spectra would be subject to opposite Doppler shift throughout the binary orbit.

\subsection{Synergy of multimessenger observations}\label{sec:discussion-multimessenger}

X-ray spectroscopy is a powerful method for deep probing of the innermost accretion flows around SMBHs, and by extension, binaries. It may in principle be used to assess the credibility of a binary candidate \citepalias{malewicz2025}, and to place bounds on some source parameters. 
Although reliable parametrized fitting will, in most cases, require the spectral resolution and effective area of a future mission like \textit{NewAthena }\citep{cruise2025}, current and archival data from \textit{XMM-Newton} \citep{jansen2001}, \textit{NuSTAR} \citep{harrison2013} and/or \textit{XRISM}  \citep{tashiro2022} can still be informative and can be used as the first steps in building the case for further observations. 
This work suggests that some preliminary estimates of binary parameters may be accessible with moderate spectral resolution (enough to identify the \added{relativistic} \feka's peak or track its centroid).

Constraints on binary parameters in the composite X-ray reflection spectrum may then be used as priors in joint-likelihood parameter estimation efforts to identify and characterize a SMBH binary from GW data \citep[as shown by][for PTA datasets]{charisi2025}, and potentially to help disambiguate time-domain observations.
Joint parameter estimation will be a collaborative effort, and it is worth considering how the diagnostic benchmarks outlined in this work may fit within a multimessenger context. 
The SMBH binaries detectable by either LISA or the PTAs are drawn from distinct populations, with different observational characteristics (luminosity, redshift, variability timescales), motivating separate approaches for each.

%
\paragraph{PTA binaries} We note in \citetalias{malewicz2025} that the main properties of a PTA binary make it an ideal target for time-resolved, high signal-to-noise X-ray spectroscopy: they are massive ($\gtrsim 10^8 \Msun$) and close-by ($z\lesssim 2$), making them bright sources with orbital periods on the order of a few months to a few decades.
They are also long-lived: the estimated time to merger for a typical PTA source is $\sim \mathcal{O}(10^4)$ years \citep[e.g.,][]{sesana2012}.
For that reason, targeted searches for individual binaries in the PTA datasets are generally (though not exclusively) prioritizing single-frequency continuous wave signals, which may have an easier time rising above the stochastic gravitational wave background \citep[e.g.,][]{schult2025}. The minimal chirping ($\dot{f}$) makes mass ratio difficult to diagnose due to the high level of degeneracy between the chirp mass $\mathcal{M}_c$ (a function of $M_{\rm tot}$ and $q$) and luminosity distance $d_L$. 
However, this problem is partly resolvable through the identification of an EM counterpart and associated host galaxy, which can be used to measure the redshift and luminosity distance (assuming a specific cosmological model). Observing a \feka\ centroid shift would help narrow down mass ratio estimates.
The inclination of the binary orbital plane 
affects the fractional GW power observed in different polarization modes, and will likely be the most difficult parameter to extract from a PTA detection of a continuous wave \citep{petrov2025} -- which, conversely, could be accessed by an X-ray spectrum.
Finally, given that spin and mass ratio are only encoded as post-Newtonian, higher-order effects in the GW signal, they will realistically only be constrainable for louder, more massive and/or higher frequency binaries \citep[e.g.,][]{corbincornish2010, mingarelli2012}.
This means that X-ray reflection constitutes the only avenue for probing the spins of most PTA SMBH binaries.


\paragraph{LISA binaries} The LISA class of SMBH binaries presents a series of observational challenges for deep X-ray spectroscopy.
On one hand, contrary to the continuous GWs targeted by PTAs, the inspiraling waveforms captured by LISA will contain a wealth of information on the masses of the two constituent BHs and the binary effective spin \citep{amaro-seoane2023}.
However, not all merger events captured by LISA will be ideal targets for deep X-ray spectroscopy.
The preferred LISA candidate for multimessenger followup would be nearby ($z\lesssim 0.1$) and as luminous (or massive) as possible.
A promising target for parameter estimation using X-ray spectroscopy would be a LISA precursor source, detected early in its inspiral -- before the binary orbital torques disrupt the mini-disks. 
As noted in \citetalias{malewicz2025}, the standard exposure time for a reflection spectrum of about 100~ks is comparable to the orbital period of a typical LISA binary, resulting in Doppler smearing of sharp reflection features. 
Early detection may mitigate those effects.
A more luminous source would also reduce the required exposure time, minimizing both Doppler smearing and orbital evolution during the observations.
Despite those caveats, the concurrent operation of LISA and NewAthena would offer unprecedented opportunities for multimessenger constraints on SMBH masses, spins and the late-inspiral accretion environment. 
This can also be mitigated by improved instrument specifications, like higher collecting areas (such as \textit{XMM-Newton}, \textit{NewAthena} or \textit{Lynx}) which scales linearly with photon count rate. 
Coordinated multi-instrument campaigns also often serve as a way to artificially increase the effective area through joint-fitting of simultaneous observations \citep[e.g.,][]{wilkins2026,moutard2023}.
An unambiguous detection of a binary SMBH may provide enough justification for an observational campaign of that caliber.

%
Though we have not explored the possibility that the binary orbits may be eccentric, Figure \ref{fig:mass_ratio_5_panels}'s centroid shift time series could conceivably deviate from a sinusoidal curve if the orbit of the brighter SMBH driving the shift is not circular \citep[e.g.,][]{hu2020}. 
GWs slowly but efficiently carry away eccentricity over a binary's evolution \citep{peters1963, peters1964}, and so while isolated LISA binaries could have largely circularized, dynamic interactions with the circumbinary disk may impart some residual eccentricity to the binary at the time of decoupling \citep{siwek2023}. As a result, a significant fraction of PTA binaries may retain some non-negligible degree of orbital eccentricity \citep[e.g.,][]{sesana2012}.
We leave this line of inquiry open for future work.

\subsection{Idealizations in the model}\label{sec:discussion-limitations}

We make a number of simplifying assumptions to make the problem tractable and it is worth discussing the implications of such simplifications. The model limitations brought up in the Discussion section in \citetalias{malewicz2025} are still broadly applicable. This includes both uncertainties inherent to single SMBH reflection modeling and simplifications in the binary model. 

\begin{figure*}
    \centering
    \includegraphics[width=\linewidth]{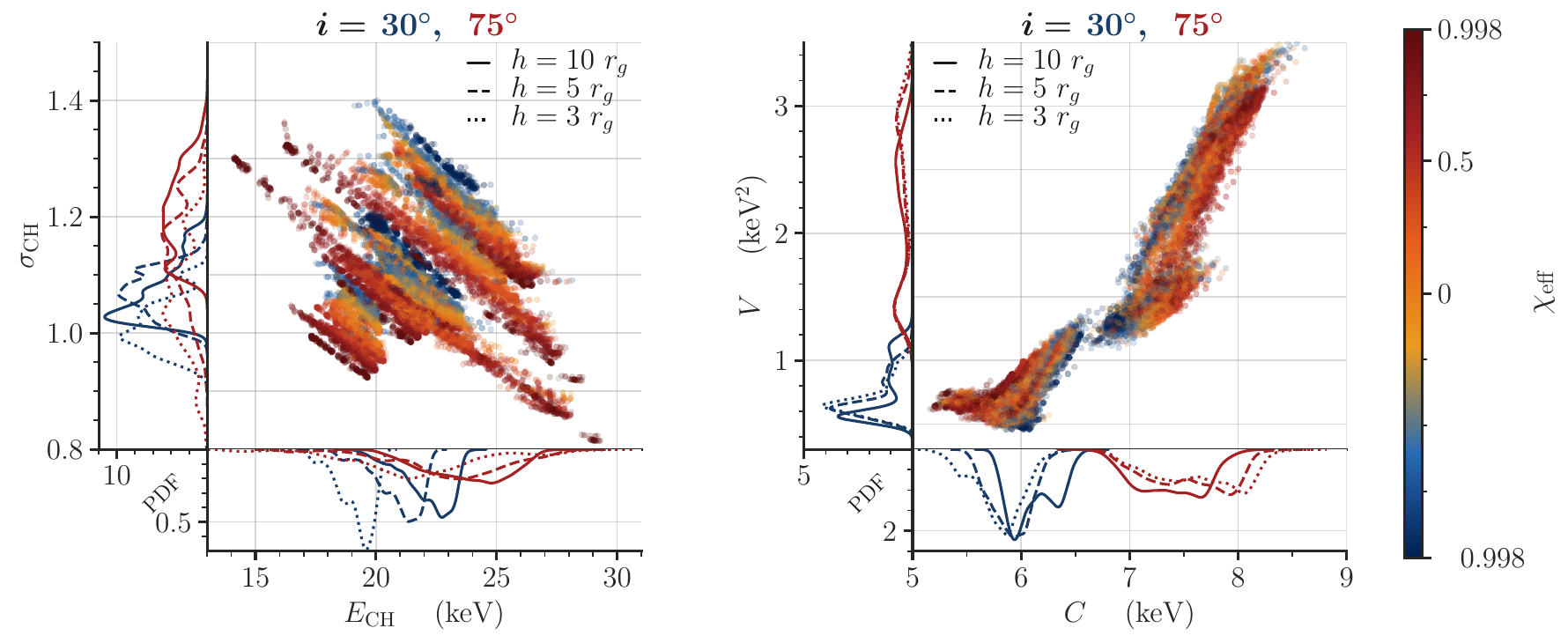}
    \caption{{\it Left:} As in Fig.~\ref{fig:compton_spin} (CH peak ${E_\textrm{CH}}$ vs. log-width ${\sigma_\textrm{CH}}$), but including three different values of coronal height, $h_c=3, 5$ and $10\,\rg$. {\it Right:} As in Fig.~\ref{fig:feka_centroid_vs_var} (\feka\ variance ${V}$ vs. centroid ${C}$), but for the same three values of $h_c$ as above. Both are plotted for inclinations $i=30\degree$ and $75\degree$.}
    \label{fig:discussion_hc}
\end{figure*}

Though our measurement approach is model-independent, the forward modeling of the individual SMBH spectra is not.
For example, we have not considered in this work how different reflection models may challenge our diagnostic approach.
There remains significant uncertainty about the nature and geometry of the corona -- though it is generally considered to be compact \citep[e.g.,][]{chen2011}.
The lamppost is a mathematical idealization for ease of computing, whose effective height serves as proxy for compactness. It reproduces observations of relativistic reflection quite well.
A lower coronal height, $h_c$, increases the innermost disk's emissivity relative to the rest of the disk and results in more marked relativistic distortions in the reflection spectrum \citep[e.g.,][]{fukumura2007, dauser2013}. For that reason, uncertainties around $h_c$ will be strongly degenerate with spin. Conversely, a higher coronal height would instead decrease the inner disk emissivity
in favor of more distant, less relativistically impacted regions. This would severely hamper our ability to measure spin, which primarily relies on detecting strong gravitational redshift from disk radii $\lesssim 6~\rg$. In that sense, our dataset errs on the side of being more conservative: the effective coronal height in this work is fixed to $10~\rg$, which is usually considered to be an upper limit of best-fit coronal heights obtained from fitting, and results in a flatter emissivity profile.

We nevertheless produced two additional populations with $h_c=3~\rg$ and $5~\rg$ to test the impact of this parameter on our results.
Figures \ref{fig:inclination} (peak and inclination) and \ref{fig:mass_ratio_5_panels} (centroid shift and mass ratio) are unchanged, even when considering all three $h_c$ values at once.
We show in Fig.~\ref{fig:discussion_hc} the impact of including these sub-populations on the right panels of Figs.~\ref{fig:compton_spin} and \ref{fig:feka_centroid_vs_var}.
The Compton hump morphology is dependent on the ionization structure of the disk, and
the height of the coronal illuminator is only one of the parameters impacting it. 
Relaxing assumptions about $h_c$ therefore predictably causes degeneracies in metric space with spin and inclination. In the left panel of Fig.~\ref{fig:discussion_hc}, which includes models for all three choices of $h_c$, the direction of the effective spin dependence in metric space is the same as in Fig.~\ref{fig:compton_spin} (right panel), but the sub-populations of points with $h_c=3~\rg$ and $5~\rg$ fill in the space between the two inclination regimes ($i=30\degree$ and $70\degree$).
This shows that without constraints on coronal height, the model-independent CH metrics cannot be used on their own to evaluate spin, but the fact that the correlation between CH morphology and binary effective spin holds regardless of $h_c$ is encouraging.

Conversely, the \feka's diagnostic power comes from its relativistic deformation, and so while the height of the corona will alter the illumination pattern (and subsequently, the disk's emissivity profile), it does not impact the spacetime effects imprinted on the fluoresced disk photons. The observations made in \S~\ref{sec:effective-spin-feka} are still broadly true: just like in the single BH case,
the dominant spin effects on the \feka's centroid $C$ and variance $V$ depend on inclination:
at low inclinations (e.g. $i=30\degree$) and high \chieff, the extended red tail to the \feka\ profile biases the centroid to low energies. In contrast, high inclinations and high \chieff\ enhance the \feka's blue wing via strong Doppler beaming and blue-shifting. As a result, the right panel of Fig.~\ref{fig:discussion_hc} is very similar to that in the right panel of Fig.~\ref{fig:compton_spin} and relatively unaffected by the different choices of $h_c$.
In summary, any quantitative spin estimation is inherently vulnerable to degeneracy, especially when coronal height is not known \textit{a priori} -- even more so than in the single BH case. Still, we find correlations between the binary's effective spin and morphological characteristics of both the \feka\ profile and the CH that persist across $h_c$ values. 
The \feka\ profile remains the most reliable carrier of spin information, but the CH supplies complementary information that may help a broadband parametrized fitting scheme break degeneracies -- especially if independent constraints on coronal height can be gleaned from other techniques, like reverberation mapping \citep[e.g.,][]{alston2020}.
Investigation into how other coronal geometries \citep[see, e.g.][]{kim2024, nekrasov2025} would impact these results are outside the scope of this study, but will need to be evaluated during the development of a binary spectral model for parametrized fitting.

Additionally, our model only considers the relativistic reflection spectra from the two mini-disks. In reality, some X-ray emission is likely to originate from the circumbinary disk, as well as from the optical broad line region ($\sim$0.01 pc) and dusty torus \citep[$\sim$ 1 pc; see][for recent examples]{xrism2024,bogensberger2025,miller2025a, miller2025b,juranova2025,li2026}.  
Spectral lines (such as the one from \feka) that are contributed by these emission regions on larger scales would be narrower and appear superposed with the relativistically broadened components contributed by the mini-disks.
This may complicate the measurement of the peak of the relativistic \feka\ profile, as some measurement uncertainty would inevitably be introduced in the disentangling of the components. On the other hand, since the narrow \feka\ components do not necessarily undergo the same periodic Doppler shifting as the relativistic mini-disks, they could, in principle, be used to define the rest frame of the circumbinary disk in multi-epoch spectra. 

We also assume no galactic or local absorption or obscuration, and no contribution from a warm absorber, or a warm corona. The contribution from these components would primarily affect the X-ray spectrum at energies $\la$ 5 keV. For the \feka\ profile to be affected, the column density of the obscuration would need to be substantial ($N_H \gtrsim 10^{23}$ cm$^{-2}$).
Outflows along the line-of-sight may also affect the \feka\ profile, mainly through complex absorption, which is common in AGN \citep{reynolds1997,brenneman2013}. They may mimic the Doppler shift expected from orbiting mini-disks, making the interpretation of a single-epoch spectrum possibly ambiguous. Multi-epoch comparisons should help address any such uncertainty.

Another potential source of degeneracy in X-ray reflection spectroscopy is the iron abundance relative to solar, $A_{\rm Fe}$, which influences both the relative strength of the Fe K$\alpha$ line profile and the depth of the photoelectric absorption edge bounding the CH. Because our results make use of model-independent morphological measurements of continuum-subtracted and peak-normalized line profiles, the \feka\ metrics depend on the shape of the line rather than absolute strength.
Super-solar $A_{\rm Fe}$ values are a common finding in X-ray reflection fits to both supermassive and stellar-mass BHs \citep[e.g.,][]{garcia2018}.
It is possible that the small changes in line shape caused by a non-solar $A_{\rm Fe}$ could affect the observed relationships between the morphological parameters presented in \S~\ref{sec:results}.

To test the robustness of our framework against variations in iron abundance, we processed three additional synthetic binary populations with $A_{\rm Fe}=0.5, 2$ and 5 through our pipeline, without changing anything about the continuum subtraction methodology. We find that the core Fe K line metrics (peak $P$ and centroid $C$) remain highly stable, shifting by less than 1\% on average. The variance $V$ shifts by 1-8\%, with $A_{\rm Fe}=0.5$ exhibiting slightly larger dispersion due to the reduced contrast between line and continuum. Compton hump metrics ($E_{CH}$ and $\sigma_{CH}$) show modest shifts in their mean values per inclination ($3-8\%$ and $10-20\%$ respectively for $A_{\rm Fe}\leq2$), though we note that the $i=75\degree$ sub-population at $A_{\rm Fe}=5$ is unable to converge to a satisfactory log-Gaussian fit to the Compton hump.
Overall, while variations in $A_{\rm Fe}$ introduce additional scatter, the reported morphological trends and correlations remain robust.

Like in \citetalias{malewicz2025}, we assume that the binary orbital axis, the mini-disk rotation axis, and the SMBH spins are all parallel. The assumption of coplanarity means that they all share the same inclination angle $i$.
\citet{yu2001} for example considered the case where the mini-disks are misaligned with respect to the binary orbital plane and one another \citep[see also][]{jovanovic2014,jovanovic2020}.
They computed the resulting multi-peaked \feka\ profile and found that, if the mini-disk luminosities are comparable, it may be possible to identify two distinct blue peaks and thus estimate the mini-disk inclinations.
Our approach differs, since we calculate the reflection continuum by default, and not just the \feka\ profiles.
It may be interesting to allow for a low to moderate ($|\Delta i|< 20\degree$) inclination difference between the two mini-disks to reassess the conclusions of \citet{yu2001}. 
This may however increase the likelihood of cross-illumination of one mini-disk by its companion's corona, especially at lower orbital separations. 
Furthermore, a misalignment between the angular momentum of the mini-disk gas and the SMBH spin can result in precession and warping \citep[e.g.,][]{bardeen1975,fragile2026}, which we do not consider in our model.

%

We focus on the regime where the binary can reasonably be treated as two mini-disks, each responsible for their own reflection spectrum.
However, previous studies have shown that the composite \feka\ profile of SMBH binary remains a crucial diagnostic tool in alternative regimes \citep{mckernan2013,nagele2026c}. At smaller orbital separations ($\lesssim 40~r_g$), the mini-disk surfaces are reduced, while the CBD and accretion streams seep closer to the two SMBHs. Their contribution to the X-ray reflection spectrum can no longer be assumed to be negligible, and neither can the possibility of cross-illumination of one mini-disk by its companion's corona.
While the approach taken in this study may no longer hold, GRMHD (general relativistic magnetohydrodynamic) simulations with radiative-transfer can provide useful insights into the impact of non-axisymmetric gas density profiles and rapidly evolving spacetimes on \feka\ fluorescence. Notably, \citet{nagele2026c} finds that at separations of $20~r_g$, \feka\ emission from the binary accretion streams shows a distinct periodicity associated with an over-density at the inner edge of the CBD, at $\sim 0.1-0.2\times$ the binary orbital period.
Another binary regime we have not considered is that of small mass ratios ($q\sim10^{-3}$), but the \feka\ profile remains informative even then. \citet{mckernan2013} made observational predictions of LISA- and PTA-mass sources where the inward migration of the much lighter secondary SMBH through the accretion disk of the primary carves out low-density rings at lower and lower radii, resulting in characteristic ripples in the \feka\ profile.

Due to accretion inversion described in \S \ref{sec:methods-compute}, a non-negligible fraction of observable binary systems is likely to host a secondary SMBH accreting at a super-Eddington rate (see, e.g., Figure 1 of \citetalias{malewicz2025}), which could greatly affect its X-ray emission. 
In the low-accretion limit, accretion inversion may have the opposite effect: the primary SMBH can become highly sub-Eddington and radiatively inefficient \citep[see][]{tiede2025}. 
In either case, it is possible that only one of the two mini-disks may be in the thin disk regime and thus dominate the reflection features of the composite.
In cases where the coronal irradiation is too weak or too harsh, its mini-disk may be under- or over-ionized and thus unable to produce salient atomic reflection features \citep[e.g.,][]{garcia2010}.
This makes for an interesting limiting case, where the binary observational challenge is collapsed back into single SMBH reflection, with some possible attenuated contribution from its companion. If so, properties like spin and inclination may then be accessible through standard, single SMBH parametrized fitting.

If the reflection from the super-Eddington accreting secondary mini-disk is suppressed, the composite spectrum will mainly retain the imprint of the primary, even though X-ray emission from optically thick outflows \citep[e.g.,][]{king2003,zhang2024} around the secondary could complicate this picture.
Unequal mass ratios also imply minimal orbital motion from the primary SMBH, weakening the diagnostic power of spectral metrics caused by the orbital Doppler shift ($\Delta C$, for example). %
Alternatively, if the secondary SMBH dominates the reflection spectrum and the X-ray emission from the primary mini-disk is suppressed, multi-epoch measurements of the \feka\ centroid will show great variability, as was illustrated in the $q=0.2$ panel of Figure \ref{fig:mass_ratio_5_panels}.

Finally, while there is broad consensus about the phenomenon of accretion inversion, there is no unique prescription for $\kappa(q)\equiv \dot{M}_2/\dot{M}_1$, the ratio of the accreted gas between the primary and the secondary SMBH (see Equations \ref{eq:lambda1} and \ref{eq:lambda2}).
We use $\kappa$ as derived by \citet{kelley2019}, based on the 2D-hydrodynamical simulations of \citet{farris2014}, which predicts a maximum inversion factor of $\kappa_{\rm max} \approx 21$ for $q \approx 0.1$. 
The lowest mass ratio we consider in this study is $q=0.2$, where the inversion factor is far less pronounced ($\kappa(0.2)\simeq3.3$).
Still, we find that a more conservative parametrization of the same fitting formula, where the maximum inversion factor is decreased to match more recent simulations that indicate relative rates $\kappa_{\rm max}$ as low as about 2 at $q=0.1$ \citep[e.g.,][]{tiwari2025,siwek2022,dittmann2021,duffell2020,munoz2020}, 
did not alter the conclusions of this paper. It slightly lowered the number of $q=0.2$ binaries where $\lambda_2\geq 2$ (from 12 to 10\% of all binaries in our sample).


\section{Conclusions}\label{sec:conclusion}

We use a population-level statistical approach to explore the diagnostic capabilities of the composite X-ray reflection spectrum from two mini-disks in an SMBH binary. This is a follow-up to our work in \citetalias{malewicz2025}, where we investigated the feasibility of identifying binaries through their X-ray reflection spectra.
We develop model-independent metrics to assess the shape of the prominent spectral components of the X-ray reflection spectrum and search for their correlation with the underlying SMBH binary parameters. This approach is an alternative to the commonly used full parametrized fitting of the X-ray spectrum, which has not yet been developed in the case of binary SMBHs. Our measurements indicate that the relativistic \feka\ line and the Compton reflection hump of an SMBH binary can be correlated with binary parameters.
These metrics can be used to place preliminary constraints on source properties by capturing key binary effects before a more detailed spectral fitting approach is available. We find that:

\begin{itemize}
\item The location of the blue peak in the relativistic \feka\ profile is a relatively robust diagnostic of the inclination angle of the observer relative to the binary orbital plane (\S~\ref{sec:inclination}).
This correspondence arises because the inclination angle determines the line-of-sight velocity of the disk gas and the Doppler shift undergone by emitted line photons, which in turn determines the location of the profile peak. Because the inclination is, in principle, a single-epoch measurement (i.e., it is based on a single-epoch X-ray spectrum), this is a binary parameter that should be pursued first and could be used as a prior in joint-likelihood parameter estimation efforts with GW data.

\item Given a few epochs of spectra, one may be able to determine the orbital phase of the binary and place constraints on its mass ratio by monitoring the orbital Doppler shifting of the \feka\ centroid (\S~\ref{sec:mass-ratio}).
In a circular binary, the \feka\ centroid's shifting is sinusoidal and largely driven by the orbital motion of the secondary SMBH, which accretes at a higher rate and tends to be brighter.
\item Of all binary parameters, SMBH spin effects are most subtle and prone to measurement degeneracies. The spin configuration of a binary SMBH does, however, leave measurable imprints on the composite spectrum -- most notably in the relativistic \feka\ profile (\S~\ref{sec:effective-spin-feka} and \ref{sec:individual-spin}) and the Compton hump (\S~\ref{sec:effective-spin-CH}). 
Specifically, using our model-independent metrics, we find that despite the degeneracies, it may be possible to assess whether a binary SMBH contains a highly-spinning SMBH by searching for the distinctive spectral signatures of ionized emission and absorption very near to the ISCO -- where relativistic effects are most pronounced.
\end{itemize}

These diagnostics are most powerful in multimessenger campaigns, where they can complement, confirm and contextualize GW measurements (\S~\ref{sec:discussion-multimessenger}).
\begin{itemize}
\item PTA-targeted SMBH binaries -- massive, nearby and at orbital periods of months to decades -- are ideal for time-resolved, multi-epoch X-ray spectral monitoring. In this regime, GWs constrain the orbital phase and frequency with increasingly high precision, but chirp mass, inclination, and luminosity distance remain harder to disentangle.
X-ray spectroscopy offers natural complementarity: the relativistic \feka\ profile encodes the inclination of the orbital plane (assumed co-planar with the mini-disks), while multi-epoch centroid tracking can constrain the mass ratio.
In the case of a compelling but unconfirmed binary candidate, X-ray measurements could inform priors on orbital frequency and phase to help the GW signal rise above the stochastic background.
Markers of high SMBH spin may also be imprinted on and recoverable from the composite reflection spectrum, offering a rare window into the spin configuration of the binary during its early inspiral.

\item Conversely, the LISA detectability window -- late inspiral and merger of lower-mass SMBH binaries, potentially at considerably higher redshifts -- poses greater observational challenges.
Still, LISA precursor sources detected early in their inspiral make compelling multimessenger targets: LISA will deliver exquisite constraints on the SMBH masses and effective spin, while X-ray spectroscopy can independently constrain mass ratio and inclination.

\end{itemize}

Overall, we find that X-ray reflection spectroscopy of SMBH binaries has the potential to provide valuable constraints on binary parameters, and should be an important component of multi-messenger observations during the upcoming era of low-frequency gravitational waves.

\begin{acknowledgements}

J.M., T.B., D.R.B., T.D. acknowledge support from the NSF grant
AST-2307278 
and L.B. from the NSF grant
AST-2307279. 
D.R.B. is also supported from NASA award
80NSSC24K0212 and NSF grant AST-2407658. 
T.D. acknowledges support from the DFG research unit FOR 5195 (project number 443220636, grant number WI 1860/20-1).
\end{acknowledgements}


\bibliography{sample701}
\bibliographystyle{aasjournalv7}
\end{document}

%% file: def.tex
\usepackage{bm} 
\usepackage[dvipsnames]{xcolor}
\usepackage{pifont}
\usepackage{amsmath}
\usepackage{soul} 

\defcitealias{malewicz2025}{Paper I}
\defcitealias{tiwari2025}{T25}
\defcitealias{kelley2019}{K19}

\definecolor{square}{HTML}{E86B1E}
\definecolor{cross}{HTML}{B42F23}
\definecolor{circle}{HTML}{5C0B0C}
\definecolor{plus}{HTML}{5A778C}
\definecolor{star}{HTML}{E87C1E}

\newcommand{\Plus}{\mathord{\text{\ding{58}}}}
\newcommand{\Cross}{\mathord{\text{\ding{54}}}}
\newcommand{\Star}{\mathord{\text{\ding{72}}}}
\newcommand{\Circle}{\mathord{\text{\ding{108}}}}

\newcommand{\relxill}{\texttt{relxill}}
\newcommand{\xspec}{\texttt{Xspec}}
\newcommand{\feka}{Fe\,K$\alpha$}
\newcommand{\ltot}{\ensuremath{\lambda_{\mathrm{tot}}}}
\newcommand{\degree}{\ensuremath{^{\circ}}}
\newcommand{\chieff}{\ensuremath{\chi_{\mathrm{eff}}}}
\newcommand{\Msun}{\ensuremath{M_{\odot}}}

\newcommand{\rin}[1][]{%
  \ifx\relax#1\relax
    \ensuremath{r_{\mathrm{in}}}%
  \else
    \ensuremath{r_{{\mathrm{in}, #1}}}%
  \fi
}

\newcommand{\rout}[1][]{%
  \ifx\relax#1\relax
    \ensuremath{r_{\mathrm{out}}}%
  \else
    \ensuremath{r_{{\mathrm{out}, #1}}}%
  \fi
}

\newcommand{\rg}[1][]{%
  \ifx\relax#1\relax
    \ensuremath{r_{g}}%
  \else
    \ensuremath{r_{g, #1}}%
  \fi
}

\newcommand{\risco}[1][]{%
  \ifx\relax#1\relax
    \ensuremath{r_{\mathrm{ISCO}}}%
  \else
    \ensuremath{r_{#1{\mathrm{(ISCO)}}}}%
  \fi
}

\newcommand{\rl}[1][]{%
  \ifx\relax#1\relax
    \ensuremath{r_{\mathrm{L}}}%
  \else
    \ensuremath{r_{#1{\mathrm{(L)}}}}%
  \fi
}